# Random packing fraction of binary similar particles: Onsager's excluded volume model revisited[1]

H.J.H. Brouwers

Department of the Built Environment, Eindhoven University of Technology, P.O. Box 513, 5600 MB Eindhoven, the Netherlands

[1] This paper is dedicated to my late father J.H. (Jef) Brouwers (Heer, 12 January 1916 – Maastricht, 17 February 2019).

**Abstract**

In this paper, the binary random packing fraction of similar particles with size ratios ranging from unity to well over 2 is studied. The classic excluded volume model for spherocylinders and cylinders proposed by Onsager [1] is revisited to derive an asymptotically correct expression for these binary packings. From a Taylor series expansion, it follows that the packing fraction increase by binary polydispersity equals $2f(1 - f)X_1(1 - X_1)(u - 1)^2 + O((u - 1)^3)$, where f is the monosized packing fraction, $X_1$ is the number fraction of a component, and u is the size ratio of the two particles. This equation is in excellent agreement with the semi-empirical expression provided by Mangelsdorf and Washington [2] for random close packing (RCP) of spheres. Combining both approaches, a generic explicit equation for the bidisperse packing fraction is proposed, which is applicable to size ratios well above 2. This expression is extensively compared with computer simulations of the random close packing of binary spherocylinder packings, spheres included, and random loose sphere packings ($1 \leq u \leq 2$). The derived generic closed-form and parameter-free equation, which contains a monosized packing fraction, size ratio, and composition of particle mix, appears to be in excellent agreement with the collection of computer-generated packing data using four different computer algorithms and RCP and random loose packing (RLP) compaction states. Furthermore, the present analysis yields a monodisperse packing fraction map of a wide collection of particle types in various compaction states. The explicit RCP-RLP boundaries of this map appear to be in good agreement with all reviewed data. Appendix A presents a review of published monodisperse packing fractions of (sphero)cylinders for aspect ratios l/d from zero to infinity and in RLP and RCP packing configurations, and they are related to Onsager's excluded volume-based model. Appendix B presents a derivation of the binary packing fraction of disks in a plane ($\mathbb{R}^2$) and hyperspheres in $\mathbb{R}^D$ ($D > 3$) with a small size difference, again using this model.

## 1. Introduction

The packing of granular matter is an old physical, biological, and mathematical puzzle that received much attention in past millennia [3]. The packing fraction of particles is the simplest example and is also one of the most basic features of a granular matter system. Packing problems are ubiquitous and arise in the transportation, agricultural, packaging, and communication industries. The optimal packing of hard spheres in higher dimensions is of interest in error-correcting codes in communication theory [4]. Furthermore, attention has been paid to revealing the packing geometries and the route to understanding liquids, glasses, metals and metalloids, and granular materials. The packing of hard spheres is the simplest example of such systems, whether crystalline or amorphous. Besides being of scientific importance, the packing fraction of particles is also technically of relevance. Examples are concrete [5], ceramics [6], catalyst beds [7], coal-ore stackings [7, 8], composite materials [9, 10], fibrous filters [11], clothing materials, nonwoven materials and fibrous hygienic items [12]. Their technical properties depend to a considerable extent on the closeness of the packing of the particles.

The densest packing of equal particles is obtained for ordered (crystalline) arrangements or lattice packing, which can be computed. For example, for spheres, the dc (diamond cubic), sc (simple cubic), bcc, and fcc/hcp lattices have packing fractions of $3^{1/2}\pi/16$ (≈ 0.34), $\pi/6$ (≈ 0.52), $3^{1/2}\pi/8$ (≈ 0.68), and $2^{1/2}\pi/6$ (≈ 0.74), respectively. For a disordered (random/amorphous) particle packing, the packing fraction depends on the densification state (*e.g.*, loose and dense). Numerous numerical and experimental studies have confirmed a common upper limit for the random close packing (RCP) value of disordered frictionless sphere packings, $f_{spe}^{rcp}$, with a consistent value of approximately 0.64 [2, 6-8, 13-27]. Mathematically, the RCP state is difficult to define, because, by introducing order, higher packing factions can be obtained. In [4, 28], the "maximum random jammed" (MRJ) state was introduced as a concept, defined by configurations with minimal values of typical order parameters. For strictly jammed packing, the lowest ordering yielded a packing fraction of 0.64. For random loose packing (RLP) of monodisperse particles, a reproducible packing fraction is also found in [7, 14, 23, 29-34], for spheres, $f_{spe}^{rlp} \approx 0.54$, which is a generally accepted value for this lower packing limit. In [4, 28], it was shown that strictly jammed sphere packings may reach a packing fraction as low as 0.49, but also that this comes with higher ordering. These sphere packings are members of a whole family of particle types, each having their characteristic loosest and densest packing configuration. Members of this family include spherocylinders, cylinders (with planar ends), and cubes (and the four other platonic solids).

When uniformly shaped particles of different sizes are randomly packed, *i.e.,* generating a polydisperse packing, the packing fraction increases compared with the monosized packing, that is, the packing of congruent (or identical) particles, and the associated volume contraction depends on the particle size distribution. By combining two equally shaped (similar) particles of different sizes, such a polydisperse packing can be readily assembled. In this study, this specific polydisperse particle packing is analyzed, *viz.* the packing of two discretely sized and equally shaped particles, referred to as binary mixtures. Binary packing of similar particles was studied in [2, 6, 15, 17, 19, 21, 35-40]. For binary mixes with a large size ratio u ($u \rightarrow \infty$), *i.e.* two noninteracting fractions, an analytical expression for the binary packing fraction is available [6, 21, 39, 40]. In [35, 36], analytical equations were derived for the packing fraction of crystalline structures consisting of randomly placed bidisperse hard spheres for the other limit, *viz.* a size ratio u close to unity ($u \rightarrow 1$). It was shown that the underlying approach for studying volume distortion introduced by unequal sphere pairs is also applicable to randomly packed spheres [37], yielding an analytical expression for the binary packing fraction for this limit. Furthermore, though the binary packing of similar particles with a small size difference is a relatively simple polydisperse system, it forms the basis of the packing description of polydisperse arrangements, for instance, with geometric and binomial particle size distributions [38]. They can be statistically described by considering all binary combinations of the particle size classes present [38], so an accurate description of the packing fraction of each binary combination is crucial.

Here, the packing fraction of random binary similar particles with small size differences is revisited using Onsager's model of excluded volume. Onsager developed this original model for the isotropic liquid-to-nematic (I-N) phase transition of hard rod-like (spherocylinders and cylinders) particles, which was published in his seminal paper in 1949 [1]. This excluded volume concept was introduced by Kuhn in 1934 to study polymeric chains [41]. Onsager demonstrated that a phase transition can be predicted based on two-particle (spherocylinders or cylinders) interactions, represented by the second virial term in an expansion of the free energy of the system. The Onsager expressions for the excluded volume have mostly been used to study the packing of monosized particles, although his model allows for two (sphero)cylinders with different diameters and lengths. The polydisperse model has been used to study the phase transitions of particles of different lengths [42-44], diameters [45], or constant volumes [46].

To the best of the author's knowledge, Onsager's excluded volume model has not yet been used to study the packing of binary particles. Specifically, the packing of binary similar particles, *i.e.* particles with the same aspect ratio (length-diameter), is the topic of this study. The present paper focuses on packing fraction only, that is, the ratio of the volume of the particles to the total volume they occupy, which is the *primary* feature of a particle assembly.

The remainder of this paper is organized as follows. In Section 2, Onsager's excluded volume of monosized and binary pairs of (sphero)cylinders is revisited. Using a statistical approach, these expressions are used to derive a closed-form expression for the binary packing fraction of the aforementioned particles. It appears that the expressions for spherocylinders and cylinders are identical and do not depend on their shape (governed by aspect ratio). The variation in the monosized packing fraction by binary polydispersity is a function of the concentration product $X_1(1 - X_1)$ and is proportional to the square of the relative size difference $(u – 1)$. In Section 3, a semi-empirical expression concerning the RCP fraction of binary spheres, based on the contraction function presented by Mangelsdorf and Washington [2], is introduced. It is shown that, for $u \rightarrow 1$, this semi-empirical expression transforms into the expression obtained in Section 2, based on Onsager's excluded volume model. As Onsager's model also allows for binary packings other than spheres, this semi-empirical expression for the RCP of spheres is generalized by introducing the factor $(1 – f)$ that follows from the Onsager-based model (Section 2). The quadratic expressions obtained in $(u – 1)$ are also compared with the first-order expression originally proposed in [37]. In Section 4, both new expressions are compared with the computationally generated RCP of spherocylinders, including spheres. A comparison with the RLP of spheres, which have $(1 – f)$ which is approximately 25% larger than that of the RCP of spheres, is also presented. These comparisons yield excellent agreement over the entire compositional range and size ratios u well beyond 2. In Section 5, the simulations, both RLP and RCP, are further analyzed by scaling the results and comparing them with the present model, and illustrating the effect of the contraction functions from [1], [2], and [37]. The packing fraction increase by binary polydispersity is governed by the factor $f(1 – f)$. Based on a literature review, the magnitude of this factor for a number of particle types and packing configurations (from loose to dense) is analyzed in Section 6. The conclusions are collected in Section 7.

While this paper focuses on bidisperse packings in 3 dimensions (D = 3), Appendix A presents a review of monodisperse (sphero)cylinders packings. This appendix reveals that, for monodisperse packings, Onsager's excluded volume-based model is qualitatively correct. This is no surprise, as the model applies to colloids in thermal equilibrium. Despite this observation, this paper comes to the important conclusion that applying the model yields a parameter-free closed-form expression that predicts the *effect of bidispersity on the packing fraction* correctly. In Appendix B, the excluded volume approach is applied to $\mathbb{R}^2$ (binary disks/cylinders randomly packed in a plane) and $\mathbb{R}^D$ (D > 3, binary hyperspheres or “D-spheres”).

To summarize, by revisiting and extending Onsager’s excluded volume model to binary particle packings, a generic and accurate packing expression is obtained that is applicable to size ratios of at least 2, RLP and RCP, and a large collection of different spherocylinders. Hence, Onsager’s classical work contains much more information and has a wider application than what is generally appreciated.

## 2. Binary random packings of (sphero)cylinders

In this section, the binary packing of similar cylinders and spherocylinders is analyzed, *viz.* the packing of two discretely sized particles of equal shape, termed binary mixtures, using the theory of excluded volume by Onsager [1] developed for randomly oriented (sphero)cylinders. Zocher [47] discovered that colloidal solutions of rod-like particles undergo a phase transition when they exceed a critical concentration. They provide a good model for liquid crystals, and Onsager realized that a system of hard rods can transition from the isotropic phase to the nematic phase when the density is sufficiently increased. Furthermore, based on the insight that only two-body interactions were necessary to explain this isotropic-nematic (I-N) transition, he developed a physically and mathematically lucid model based on the volume of the two particles and their excluded volume. The results of Onsager's model are of great fundamental interest, and they have also proven to be useful for studying percolation [48, 49] and monodisperse particle packing [50, 51]. First, the packing fraction of both monodisperse spherocylinders and cylinders is recapitulated, and subsequently, for the first time, Onsager's excluded volume model is applied to the binary packing case.

### 2.1 Monodisperse packings

Here, the case of monosized (congruent) cylinders and spherocylinders is considered. Both are centrally symmetric particles with diameter d, length l, and aspect ratio l/d, and the spherocylinders are capped at both ends by hemispheres with this diameter, so that their total length is l + d. The volumes of two of these particles are given by

$$2V_{spc} = \frac{\pi d^3}{3} + \frac{\pi d^2 l}{2} \ , \tag{1}$$

and

$$2V_{cyl} = \frac{\pi d^2 l}{2} \ , \tag{2}$$

for spherocylinders and cylinders, respectively.

To assess the volume occupied by the two particles, Onsager computed the orientationally averaged excluded volume, $V_e$. This is the volume that is inaccessible to one particle because of the presence of the other and which depends on the particle size and shape. This excluded volume of a particle pair, see Figure 1, is given by [1]

$$V_{e,spc} = \frac{4\pi d^3}{3} + 2\pi d^2 l + \frac{\pi d l^2}{2} \ , \tag{3}$$

and

$$V_{e,cyl} = \frac{\pi^2 d^3}{8} + \frac{\pi(\pi+3) d^2 l}{4} + \frac{\pi d l^2}{2} \ , \tag{4}$$

for spherocylinders and cylinders, respectively. The last term in Eqs. (3) and (4) prevail when l/d is large and is identical for spherocylinders and cylinders. As was said, both equations follow from averaging the relative orientations [1]. In Figure 1, the concept of particle packing and excluded volume is graphically explained, taking binary spheres as an example. The packing fraction is the ratio of the volume of the particles to the total volume they occupy.

Onsager derived the excluded volume to assess phase transitions, which are concentration-related in liquid crystals. Therefore, following the approach of Onsager [1], here, the packing fraction of the (sphero)cylinders is defined by the particle volume and excluded volume:

$$f_{spc} = \frac{2V_{spc}}{V_{e,spc}} \ , \ f_{cyl} = \frac{2V_{cyl}}{V_{e,cyl}} \ . \tag{5}$$

The excluded volume holds for a pair of particles (Figure 1), so, in Eq. (5), two particle volumes need to be taken to assess the packing fraction. Substituting Eqs. (1) and (3) into Eq. (5) yields

$$f_{spc} = \alpha\left(\frac{1 + \frac{2\alpha}{3}}{1 + 4\alpha + \frac{8\alpha^2}{3}}\right) = \left(\frac{\alpha^{-1} + \frac{2}{3}}{\alpha^{-2} + 4\alpha^{-1} + \frac{8}{3}}\right) \tag{6}$$

for spherocylinders. Substituting Eqs. (2) and (4) into Eq. (5) yields

$$f_{cyl} = \alpha\left(1 + \frac{(\pi+3)\alpha}{2} + \frac{\pi\alpha^2}{4}\right)^{-1} = \alpha^{-1}\left(\alpha^{-2} + \frac{(\pi+3)}{2\alpha} + \frac{\pi}{4}\right)^{-1} \tag{7}$$

for cylinders, in which the following aspect ratio has been introduced:

$$\alpha = \frac{d}{l} \ . \tag{8}$$

In Appendix A, the (sphero)cylinder packing expressions, Eqs. (6) and (7), provided by using the excluded volume expressions of Onsager, are compared with experimental and computational data for a wide range of $\alpha$, both for RCP and RLP configurations.

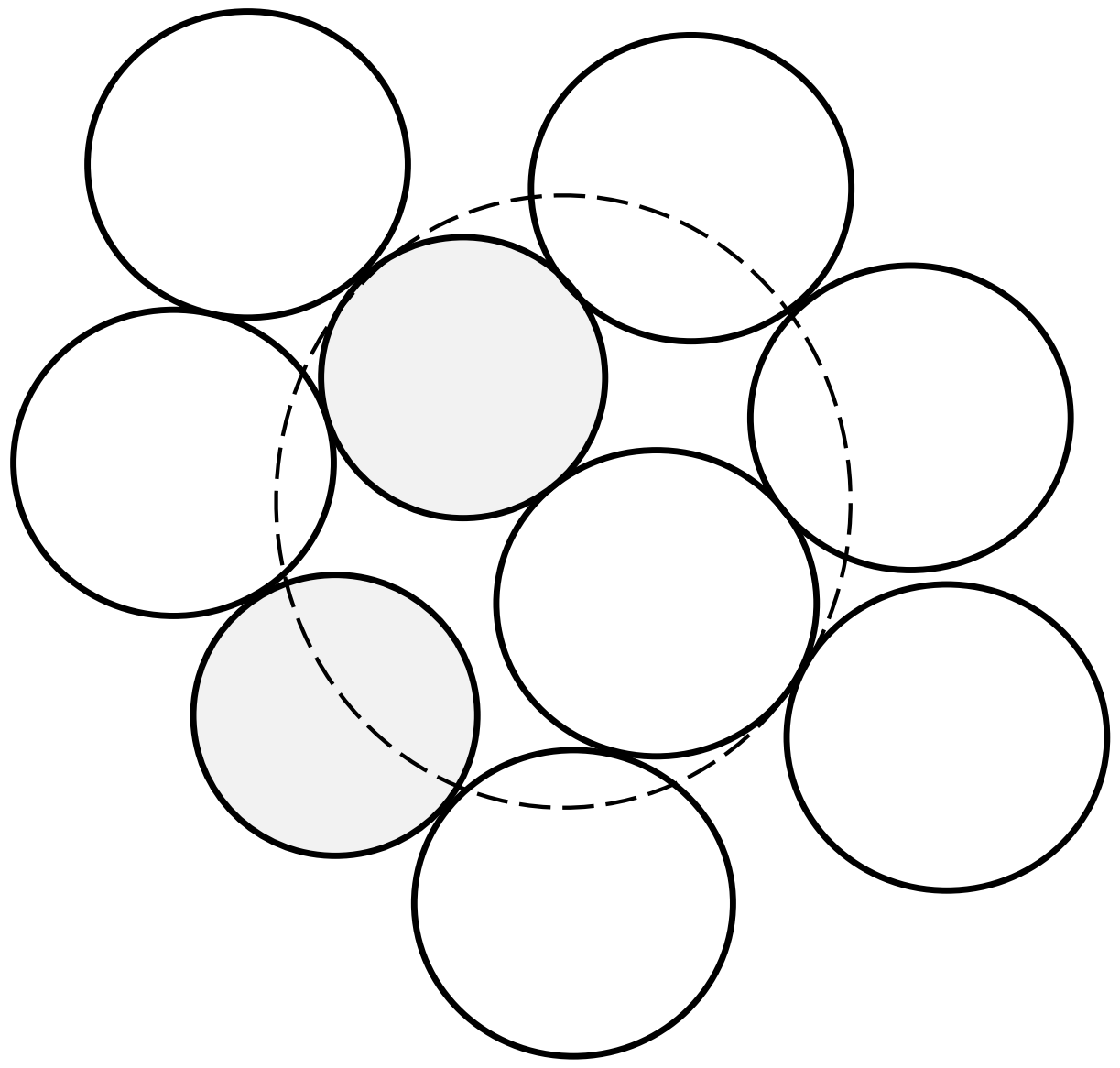

**Figure 1** Packing of bidisperse spheres; smaller spheres are shaded. Volume fraction occupied by spheres is the packing fraction (space between is the void fraction), and excluded volume of one pair of spheres is encircled by a dashed line.

It is seen that both expressions characterize the observed trends with respect to $\alpha$ quite well, and both equations appear to be in qualitative agreement with published data on monodisperse particles. With a slight modification, they are in good agreement in the entire aspect ratio range ($0 \le \alpha \le \infty$). In what follows, the effect of bidispersity on the packing fraction is quantified, again using the excluded volume theory of Onsager.

## 2.2 Particle volume of binary assemblies

Here, the case of binary packing of (discretely sized) similar cylinders and spherocylinders is considered, with a normalized number distribution

$$P(d) = X_1\, \delta(d - d_1) + X_2\, \delta(d - d_2) \quad , \tag{9}$$

where δ indicates the Dirac delta function, and $X_1$ and $X_2$ are the number fractions of the two components for which the following identities hold

$$X_1 + X_2 = 1; X_1^2 = X_1 - X_1X_2;\ X_2^2 = X_2 - X_1X_2 \quad , \tag{10}$$

where d denotes the particle diameter. As we study an assembly of two particle sizes with the same shape (but different size), all dimensions have an identical size ratio u, so

$$\frac{d_1}{d_2} = \frac{l_1}{l_2} = u;\ \frac{V_{1,spc}}{V_{2,spc}} = \frac{V_{1,cyl}}{V_{2,cyl}} = u^3 \quad . \tag{11}$$

This also implies that small and large particles also possess identical aspect ratios. The convex addition of the two spherocylinders volumes (see Eq. (1)) is

$$2V_{spc} = 2V_{1,spc}X_1 + 2V_{2,spc}X_2 = \left(\frac{\pi d_1^3}{3} + \frac{\pi d_1^2 l_1}{2}\right)X_1 + \left(\frac{\pi d_2^3}{3} + \frac{\pi d_2^2 l_2}{2}\right)X_2 =$$
$$2V_{2,spc}(u^3X_1 + X_2) = 2V_{2,spc}(1 + (u^3 - 1)X_1) \quad , \tag{12}$$

and the convex addition of the two cylinders volumes, as shown in Eq. (2), reads

$$2V_{cyl} = 2V_{1,cyl}X_1 + 2V_{2,cyl}X_2 = \frac{\pi d_1^2 l_1}{2}X_1 + \frac{\pi d_2^2 l_2}{2}X_2 = 2V_{2,cyl}(u^3X_1 + X_2) = \ ,$$
$$V_{2,cyl}(1 + (u^3 - 1)X_1) \ , \tag{13}$$

where Eqs. (10) and (11) have been applied.

### 2.3 Excluded volume of binary (sphero)cylinders

In addition to the well-known expression for the excluded volume of monosized (sphero)cylinders, Onsager also provided expressions for the orientally averaged excluded volume of two (sphero)cylinders with unequal lengths and diameters.

For spherocylinders, this equation ("Eqs. (A15)-(A16)" [1]) reads

$$V_{e,spc}^{1,2} = \frac{\pi(d_1+d_2)^3}{6} + \frac{\pi(d_1+d_2)^2(l_1+l_2)}{4} + \frac{\pi(d_1+d_2)l_1l_2}{4} =$$

$$(u+1)^3\frac{\pi d_2^3}{6}+(u+1)^3\frac{\pi d_2^2 l_2}{4}+u(u+1)\frac{\pi d_2 l_2^2}{4} \tag{14}$$

(see Eq. (11)). Figure 1 shows, as an example, the excluded volume of a binary sphere pair. For $d_1 = d_2$ and $l_1 = l_2$, and hence u = 1 and $V_{e,spc}^{1,1} = V_{e,spc}^{2,2}$, Eq. (14) can be reduced to Eq. (3) for the two monosized spherocylinders. The mean excluded volume of the assembly by randomly mixing the two particle sizes follows from the statistically probable combinations of small and large (sphero)cylinders as follows:

$$V_e = \sum_{i=1}^{2}\sum_{j=1}^{2} X_i X_j V_e^{i,j} = X_1V_e^{1,1} + X_2V_e^{2,2} - X_1X_2(V_e^{1,1}+V_e^{2,2}-V_e^{1,2}-V_e^{2,1}) \quad , \tag{15}$$

whereby Eq. (10) was used. In view of Eqs. (10) and (11), and because

$$V_e^{1,1} = u^3V_e^{2,2};\;\; V_e^{2,1} = V_e^{1,2} \quad , \tag{16}$$

Eq. (15) becomes

$$V_e = (1+(u^3-1)X_1)V_e^{2,2} - X_1(1-X_1)((u^3+1)V_e^{2,2} - 2V_e^{1,2}) \quad . \tag{17}$$

Furthermore, for spherocylinders it holds

$$((u^3+1)V_{e,spc}^{2,2} - 2V_{e,spc}^{1,2}) = \left(\pi d_2^3 + \frac{3\pi d_2^2 l_2}{2} + \frac{\pi d_2 l_2^2}{2}\right)(u+1)(u-1)^2 = (V_{e,spc}^{2,2} - 2V_{2,spc})(u+1)(u-1)^2 \tag{18}$$

(see Eqs. (12) and (14)).

For the excluded volume of two cylinders with different d and l values, the following expression ("Eq. (A14)" [1]) was provided:

$$V_{e,cyl}^{1,2} = \frac{\pi^2 d_1 d_2(d_1+d_2)}{16} + \frac{\pi(d_1^2 l_1 + d_2^2 l_2)}{4} + \frac{\pi(d_2^2 l_1 + d_1^2 l_2)}{8} + \frac{\pi^2 d_1 d_2(l_1+l_2)}{8} + \frac{\pi(d_1+d_2)l_1 l_2}{4} = u(u+1)\frac{\pi^2 d_2^3}{16} + [2(u^3+1)+u(u+1)(\pi+1)]\frac{\pi d_2^2 l_2}{8} + u(u+1)\frac{\pi d_2 l_2^2}{4} \tag{19}$$

(see Eq. (11)). It can be readily observed that Eq. (19) can be transformed into Eq. (4) for u = 1, that is, when $d_1 = d_2$ and $l_1 = l_2$. Combining Eqs. (13) and (19) yields

$$((u^3+1)V_{e,cyl}^{2,2} - 2V_{e,cyl}^{1,2}) = \left(\frac{\pi^2 d_2^3}{8} + \frac{\pi(\pi+1)d_2^2 l_2}{4} + \frac{\pi d_2 l_2^2}{2}\right)(u+1)(u-1)^2 = (V_{e,cyl}^{2,2} - 2V_{2,cyl})(u+1)(u-1)^2 \quad . \tag{20}$$

Note that the analysis of bidisperse spherocylinders and cylinders yields Eqs. (18) and (20), respectively, which are very similar.

**2.4 Packing fraction of binary (sphero)cylinders**

For convenience, by definition, $X_1$ is henceforth assigned to the number fraction of the large component $X_L$, and hence $u \geq 1$ (Eq. (11)). The binary packing fraction η follows from Eqs. (5), (12), (17), and (18) for spherocylinders, and Eqs. (5), (13), (17), and (20) for cylinders, where the nominator and denominator are divided by $V_e^{2,2}$, yielding the following binary packing fraction that is a function of composition, size ratio, and monodisperse packing fraction:

$$\eta(u, X_L) = \frac{f[1 + X_L(u^3 - 1)]}{1 + X_L(u^3 - 1) - (1 - f)X_L(1 - X_L)v(u)} \quad , \tag{21}$$

with contraction function

$$v(u) = (u + 1)(u - 1)^2 \quad . \tag{22}$$

The function $v(u) > 0$ governs the decrease in the denominator, which reflects the contraction of the packing volume. For $v(u) = 0$, the binary packing fraction would take the monosized value f (see Eq. (21)), and for $v(u) > 0$, the bidisperse packing fraction $\eta(u, X_L)$ exceeds this monodisperse value (*i.e.*, bidisperse implies that $u > 1$, $X_L \neq 0$, and $X_L \neq 1$).

Expressed in large particle volume fraction $c_L$, that is, the volume of large particles divided by the volume of large and small particles (such that $c_L + c_S = 1$), Eq, (21) reads

$$\eta(u, c_L) = \frac{f\,[c_L(1 - u^3) + u^3]}{c_L(1 - u^3) + u^3 - (1 - f)c_L(1 - c_L)v(u)} \quad , \tag{23}$$

as the number fraction $X_L$ is related to this volume fraction $c_L$ by

$$X_L = \frac{c_L}{(1 - c_L)u^3 + c_L} \quad . \tag{24}$$

Based on the equations of excluded volume by Onsager for pairs of spherocylinders and cylinders and their statistical occurrence in a two-component mix, one analytical expression for their binary packing fractions is derived that is applicable to both particle types. It follows that for spherocylinders and cylinders, irrespective of their aspect ratio $\alpha$ (d/l), an identical expression follows for the bidisperse packing fraction, *viz.* Eqn. (21) or (23), which is the reason why in this general expression $\eta$ no longer requires the subscripts "spc" or "cyl". This equation furthermore shows that the deviation of the monosized packing fraction is of the order of $(u - 1)^2$. In studying the (I-N) phase transitions of polydisperse solutions of rods with slightly different lengths, Chen [44] also found a second-order dependency on the scaled distribution width.

Furthermore, although the aspect ratio (shape) and particle type (cylinder or spherocylinder) indirectly affect the monosized packing fraction f, they are not independent factors in the binary packing expression (Eq. (21)). Eqs. (21) and (23) also reveal that the volume contraction depends on the product of large and small fractions ($c_L(1 - c_L)$ or $X_L(1 - X_L)$). In [37], it was shown that the packing fraction of bidisperse packings with a small size difference can be described by a similar model to crystalline arrangements [35, 36], yielding Eqs. (21) and (23) as well. The derivation was based on counting the number of unequal (that is, large-small particle) contacts in these packings, which resulted in the same product of large and small fractions. It is remarkable that the combination of Onsager's expressions for the excluded volume of two particles, in combination with the statistical occurrence of equal and unequal particle pairs (Eq. (15)), results in the same bidisperse packing expressions v(u) for spherocylinders and cylinders, regardless of l/d (or $\alpha$).

## 3. Comparison with semi-empirical model of RCP of spheres

In the previous section, a closed-form expression was derived for the binary packing fraction section when the size difference was small (the size ratio was close to unity). To this end, the excluded volume theory of Onsager was employed, and identical expressions were obtained for the binary cylinder and spherocylinder assemblies.

In this section, this generic equation is compared with a bidisperse packing expression, which follows from revisiting a paper in which a semi-empirical expression is proposed based on random close ball packings [2]. Balls (or spheres) are spherocylinders for which l/d = 0 (or $\alpha^{-1} = 0$) (see Eq. (8)).

### 3.1 Semi-empirical model of RCP of spheres

Binary packing experiments combining a large ball ($V_{1,spe} = 0.408$ cm$^3$) with three smaller balls ($V_{2,spe}$ = 0.260, 0.1705, or 0.1032 cm$^3$) were reported in [2]. The corresponding volume ratios $u^3$ were 1.57, 2.39, and 3.95, yielding size ratios of u = 1.16, 1.34, and 1.58, respectively. The volume occupied by the balls was divided by the total number of balls, and this was compared with the volume occupied by a ball in the monosized case. In these monodiperse packings of balls with volumes of 0.408, 0.260, 0.1705, and 0.1032 cm$^3$, the volume occupied per ball was 0.642 cm$^3$ ("$V_1$"), 0.406, 0.266, and 0.160 cm$^3$ ("$V_2$"), respectively. Because f corresponds to the ratio of $V_{1,spe}$ and "$V_1$", and of $V_{2,spe}$ and "$V_2$", their monosized packing fraction follows f ≈ 0.64, implying that the RCP conditions were met. Remarkably, they published this value in September 1960 [2], later than [8], dated May 1960, but earlier than [14], dated December 1960. Also, for composed binary mixes, the total packing volume was measured and compared to the volume which would be expected based on monosized packing values. The average volume reduction per ball was computed and denoted by $V^{XS}$, for which the following semi-empirical expression was proposed:

$$V^{xs} = 4C\frac{(V_{1,spe} - V_{2,spe})^2}{\frac{1}{2}f_{spe}^{rcp}(V_{1,spe} + V_{2,spe})}X_1X_2 \quad , \tag{25}$$

with C = 0.0195 as the fitting parameter. The factor $f_{spe}^{rcp}$ is introduced here, which is the ratio of $V_{1,spe}$ and "$V_1$" and of $V_{2,spe}$ and "$V_2$", as "$V_1$" and "$V_2$" where used in the $V^{XS}$ formula provided by [2]. Contraction function Eq. (25) illustrates that, for random binary mixtures, the volume of the packing contracts; hence, the packing fraction is higher than the monosized value.

Following the experimental interpretation described in [2], the binary packing fraction can be mathematically formulated as

$$\eta_{spe}^{rcp} = \frac{(N_1 V_{1,spe} + N_2 V_{2,spe})}{(N_1 V_{1,spe} + N_2 V_{2,spe})/f_{spe}^{rcp} - (N_1 + N_2) V^{XS}} \quad , \tag{26}$$

where $N_1$ and $N_2$ are the numbers of large and small balls in the packing, respectively, $V^{XS}$ is the volume reduction per ball, and the last term of the dominator constitutes the volume contraction. The number fraction follows from

$$X_1 = \frac{N_1}{N_1 + N_2} \tag{27}$$

when $N_1 + N_2 \rightarrow \infty$ (the infinite volume limit). The volume ratio is

$$\frac{V_{1,spe}}{V_{2,spe}} = u^3 \quad . \tag{28}$$

With Eqs. (27) and (28), Eq. (26) can be rewritten as

$$\eta_{spe}^{rcp}(u, X_L) = \frac{f_{spe}^{rcp}[1 + X_L(u^3 - 1)]}{1 + X_L(u^3 - 1) - 4CX_L(1 - X_L)\left(\frac{(u^3 - 1)^2}{\frac{1}{2}(u^3 + 1)}\right)} \quad . \tag{29}$$

Therefore, it appears that the empirical approach of [2] results in the same type of equation as that theoretically derived here (Eqs. (21)) and previously in [37]. Note, however, that Eq. (29) is applicable to larger size ratios, *i.e.* u = 1.6, and not only for u – 1 ≈ 0.

The one-page paper by Mangelsdorf and Washington [2] is quite brief and does not specify the experimental conditions that underlie Eq. (25), and the equation that relates it to the binary packing fraction (Eq. (26)) is not provided. This brevity may be the reason why this original paper seems to have fallen somewhat into oblivion [52].

### 3.2 Model comparison

Eqs. (21) and (22), based on the application of Onsager's excluded volume model to similar (sphero)cylinders and Eq. (29) for RCP of spheres, derived from [2] are very similar; they differ only in the volume contraction term, which is the last term of the denominator, in particular the contraction function v(u). It is also noteworthy that, in [37], the same type of function was derived, with a contraction function that was linear in ($u^3 - 1$). To compare Eqs. (21), (22), and (29) in the vicinity of u = 1, they are asymptotically approximated using the 3rd polynomial of the Taylor series in variable u:

$$\eta(u, X_L) = \eta(1, X_L) + \eta_u(u, X_L)\,(u - 1) + \frac{1}{2}\eta_{uu}(u, X_L)\,(u - 1)^2 + O((u - 1)^3) \quad , \tag{30}$$

yielding

$$\eta(u, X_L) = f + 2f(1 - f)X_L(1 - X_L)(u - 1)^2 + O((u - 1)^3) \quad , \tag{31}$$

and

$$\eta_{spe}^{rcp}(u, X_L) = f_{spe}^{rcp} + 36Cf_{spe}^{rcp}X_L(1 - X_L)(u - 1)^2 + O((u - 1)^3) \quad , \tag{32}$$

respectively. It follows that both equations are very similar and give a quadratic deviation in (u – 1) of the monosized packing fraction. When Eq. (31) is applied to the RCP of spheres, both equations are in quantitative agreement if

$$18C = 1 - f_{spe}^{rcp} \quad . \tag{33}$$

This equality can be verified by computing 18C (≈ 0.35) and $1 - f_{spe}^{rcp}$ (≈ 0.36), which are very close and within the experimental accuracy. Accordingly, it follows that, for bidisperse mixes with small size differences, the semi-empirical fit of [2], based on bidisperse random close-packed spheres, tends to the same theoretically derived equation based on Onsager's model for bidisperse (sphero)cylinders when u ↓ 1. Interestingly, Eq. (33) also implies that the fitting parameter C in [2] is mathematically related to ($1 - f_{spe}^{rcp}$). Furthermore, it also follows that Eqs. (29), which was originally derived for the RCP of spheres only, can therefore be generalized to Eq. (21) but with the contraction function

$$v(u) = \frac{4(u^3 - 1)^2}{9(u^3 + 1)} \quad . \tag{34}$$

This means that in Eq. (29), C is replaced by $(1 - f_{spe}^{rcp})/18$, as shown in Eq. (33). In view of Eq. (33), Eqs. (21) and (34) are transformed into Eq. (29) when $f = f_{spe}^{rcp} \approx 0.64$, *i.e.* the original equation proposed in [2], derived for the RCP of spheres. However, the combination of Eqs. (21) and (34) may also allow the application of Eq. (29) to packings other than the RCP of spheres, such as (sphero)cylinders and looser packings, which are verified in detail in the following subsections.

In [37], for contraction function v(u), the expression

$$v(u) = 4\beta(u^3 - 1)/3 = 4\beta(u^2 + u + 1)(u - 1)/3 \qquad (35)$$

was put forward, with $\beta \approx 0.20$ for the RCP of spheres [37]. At the limit of $u \to 1$, Eqs. (21), (30), and (35) yield

$$\eta(u, X_L) = f + 4\beta f(1 - f)X_L(1 - X_L)(u - 1) + O((u - 1)^2) \quad . \qquad (36)$$

The models derived from the Onsager and Mangelsdorf and Washington models yield second-order terms in $(u - 1)$ (Eqs. (31) and (32)), and the equation proposed in [37] yields an approximation of the packing fraction increase that is linear in $(u - 1)$ for $u \downarrow 1$.

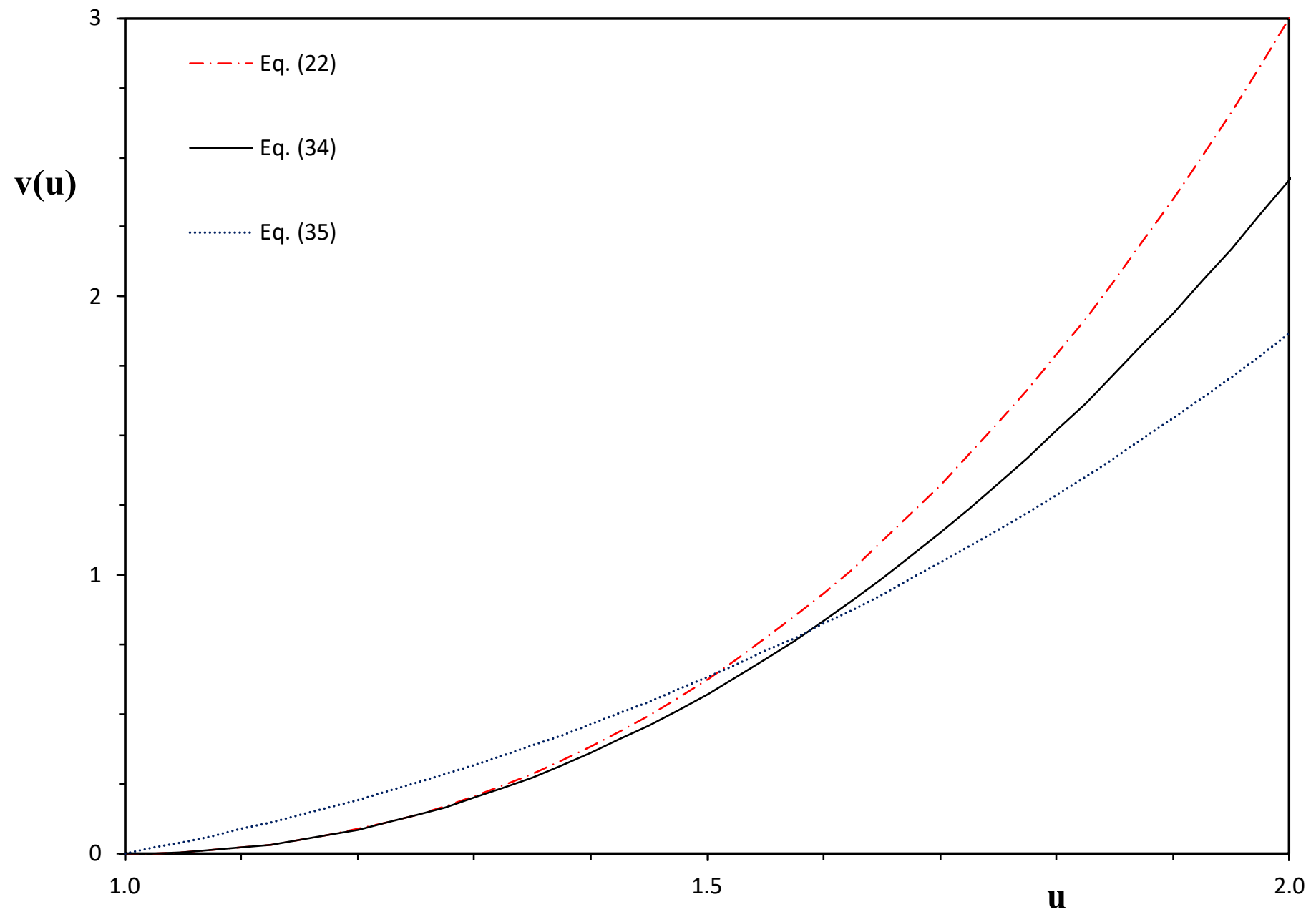

**Figure 2** Contraction functions v(u) as a function of size ratio u, u ranging from unity to two, computed using Eqs. (22), (34), and (35); in the last, $\beta = 0.20$ is employed [37].

This linear relation was already proposed in [53], in which $(u^2 + u + 1)/3$ was replaced by unity as $u \downarrow 1$. In [53], product $c_L(1 - c_L)$ was used in Eq. (36) instead of $X_L(1 - X_L)$; however, for $u \downarrow 1$, $X_L \to c_L$, see Eq. (24).

To illustrate the differences when using v(u) given by Eqs. (22), (34), and (35), v(u) is shown in Figure 2. The difference in v(u) predicted by the Onsager and Mangelsdorf and Washington models for a given u is very minor for a small (u – 1), which is not surprising, as both Eqs. (22) and (34) tend to $2(u - 1)^2$ for $u \to 1$. However, for a larger u, the difference becomes more pronounced, and at u = 2, it is approximately 25%.

Figure 2 also shows that the contraction function linear in $(u^3 - 1)$ in Eq. (35) is close to the second-order models based on the Onsager model and Mangelsdorf and Washington experiments but is less accurate in the entire range of $1 \leq u \leq 2$. For a small (u – 1), owing to its linear dependence on (u – 1), it overestimates the contraction (and packing fraction), whereas for $u \to 2$ and larger, it underestimates the volume contraction.

From the foregoing, it follows that the derivation of the Onsager theory-based model and the derivation of the Mangelsdorf and Washington data-based model yield similar equations. The former yields the introduction, in view of Eq. (33), of the term (1 – f) in Eq. (29).

To assess both models, each of which has a different contraction function v(u), *viz.* Eq. (22) and (34), in the next section, they are compared with a broad set of modern numerical packing simulations of the RCP, spherocylinders, and the RLP of spheres.

## 4. Simulated random packing fraction of bidisperse spherocylinders

In this section, Onsager- and Mangelsdorf and Washington-based models are compared to computer-generated packing of bidisperse random packings: close-packed spheres (RCP), $f_{spe}^{rcp}$ ≈ 0.64, close-packed spherocylinders (l/d > 0) with different aspect ratios, and random loose packed spheres (RLP), $f_{spe}^{rlp}$ ≈ 0.54. Note that the sphere packings are spherocylinder packings in the special case where l/d = 0. The simulation results of four different computational protocols are used.

### 4.1 Computer-generated RCP of spheres

In [37], the derived expression for the binary packing fraction linear in $(u^3 - 1)$ (Eq. (35)) was extensively compared with a broad collection of computational and experimental packing data, which were available for u = 2 only. These packing fraction values were in line with each other, but for a thorough verification, accurate binary packing fractions for u < 2 are also required. To this end, in Table 1, computer-generated data [54-56] concerning the bidisperse RCP of spheres are included for u = 1.3, 1.5, 1.7, and 2 versus the volume fraction $c_L$ of large spheres. Note that, for these larger u values, O(u – 1) ≈ 1; therefore, it is no longer close to zero. The data from [54] are generated with the same model as referred to in [20], [55] with the model of [22, 24], and [56] with the model described in [23]. The three referenced sources report slightly different monosized RCP fraction values ($f_{spe}^{rcp}$), *viz.* 0.644 [54], 0.643 [55], and 0.634 [56], which follow from $\eta_{spe}^{rcp}(u, c_L = 0)$ or $\eta_{spe}^{rcp}(u, c_L = 1)$ (Table 1). Therefore, their binary packing fraction values, $\eta_{spe}^{rcp}$, are all divided (scaled) by their monosized packing value, $f_{spe}^{rcp}$, which are included in Figure 3. First, all simulation models yield scaled binary packing fraction values that are very close to each other.

In Figure 3, Eq. (23) is set out versus $c_L$ for u = 1.3, 1.5, 1.7, and 2, using both Eqs. (22) and (34), that is, contractions following the analyses of Onsager's model and Mangelsdorf and Washington's empirical fit. For the smallest u, u = 1.3, both models are very close, as expected, in view of their identical asymptotic behavior for u ↓ 1. For larger u, in line with Figure 2, it can be seen that the Onsager contraction function results in a much larger binary packing fraction, larger than experimentally expected and simulated.

Figure 3 also confirms that Eqs. (23) and (34) can adequately predict the scaled binary RCP fraction over the entire compositional range $0 \leq c_L \leq 1$ and for the size ratio range $1 \leq u \leq 2$. This expression, a combination of the Onsager- and Mangelsdorf and Washington-based expressions, is highly accurate and useful in predicting the binary packing fraction, even up to a size ratio u of (at least) 2. This figure confirms that the use of Eq. (34) in Eq. (23) provides better agreement than Eq. (22), and this performance becomes more evident for a larger u.

a)

| $c_L$ | u = 1.3 | u = 1.5 | u = 1.7 | u = 2 |
|---|---|---|---|---|
| 0.0 | 0.6444 | 0.6444 | 0.6444 | 0.6444 |
| 0.1 | 0.6461 | 0.6478 | 0.6496 | 0.6519 |
| 0.2 | 0.6475 | 0.6510 | 0.6545 | 0.6593 |
| 0.3 | 0.6488 | 0.6538 | 0.6590 | 0.6662 |
| 0.4 | 0.6497 | 0.6561 | 0.6629 | 0.6726 |
| 0.5 | 0.6503 | 0.6578 | 0.6660 | 0.6782 |
| 0.6 | **0.6504** | **0.6586** | 0.6680 | 0.6823 |
| 0.7 | 0.6500 | 0.6582 | **0.6682** | **0.6841** |
| 0.8 | 0.6490 | 0.6563 | 0.6657 | 0.6819 |
| 0.9 | 0.6472 | 0.6520 | 0.6588 | 0.6718 |
| 1.0 | 0.6444 | 0.6444 | 0.6444 | 0.6444 |

b)

| $c_L$ | u = 1.3 | u = 1.5 | u = 1.7 | u = 2 |
|---|---|---|---|---|
| 0.0 | 0.6434 | 0.6434 | 0.6444 | 0.6444 |
| 0.1 | 0.6476 | 0.6496 | 0.6475 | 0.6523 |
| 0.2 | 0.6479 | 0.6512 | 0.6510 | 0.6608 |
| 0.3 | 0.6492 | 0.6536 | 0.6546 | 0.6678 |
| 0.4 | 0.6500 | 0.6555 | 0.6588 | 0.6748 |
| 0.5 | **0.6507** | 0.6569 | 0.6634 | 0.6803 |
| 0.6 | 0.6505 | **0.6575** | 0.6656 | 0.6834 |
| 0.7 | 0.6502 | 0.6568 | **0.6671** | **0.6839** |
| 0.8 | 0.6491 | 0.6555 | 0.6652 | 0.6814 |
| 0.9 | 0.6474 | 0.6516 | 0.6556 | 0.6711 |
| 1.0 | 0.6437 | 0.6436 | 0.6443 | 0.6445 |

c)

| $c_L$ | u = 1.3 | u = 1.5 | u = 1.7 | u = 2 |
|---|---|---|---|---|
| 0.0 | 0.6340 | 0.6340 | 0.6340 | 0.6340 |
| 0.1 | 0.6356 | 0.6372 | 0.6386 | 0.6404 |
| 0.2 | 0.6370 | 0.6401 | 0.6431 | 0.6469 |
| 0.3 | 0.6382 | 0.6423 | 0.6475 | 0.6534 |
| 0.4 | 0.6392 | 0.6453 | 0.6515 | 0.6597 |
| 0.5 | 0.6399 | 0.6472 | 0.6549 | 0.6657 |
| 0.6 | **0.6402** | **0.6484** | 0.6575 | 0.6708 |
| 0.7 | 0.6399 | **0.6484** | **0.6586** | **0.6743** |
| 0.8 | 0.6389 | 0.6468 | 0.6571 | **0.6743** |
| 0.9 | 0.6370 | 0.6426 | 0.6506 | 0.6661 |
| 1.0 | 0.6340 | 0.6340 | 0.6340 | 0.6340 |

**Table 1** Binary packing fraction $\eta_{spe}^{rcp}$ (u, $c_L$) following from computer simulations of RCP of binary spheres with 4 different size ratios u according to data from a) [54], b) [55], c) [56]. Table shows that these sources reported different monosized packing values ($f_{spe}^{rcp}$), viz. 0.644 [54], 0.643 [55], and 0.634 [56]. The maximum binary packing values $\eta_{spe,max}^{rcp}$ are indicated in bold.

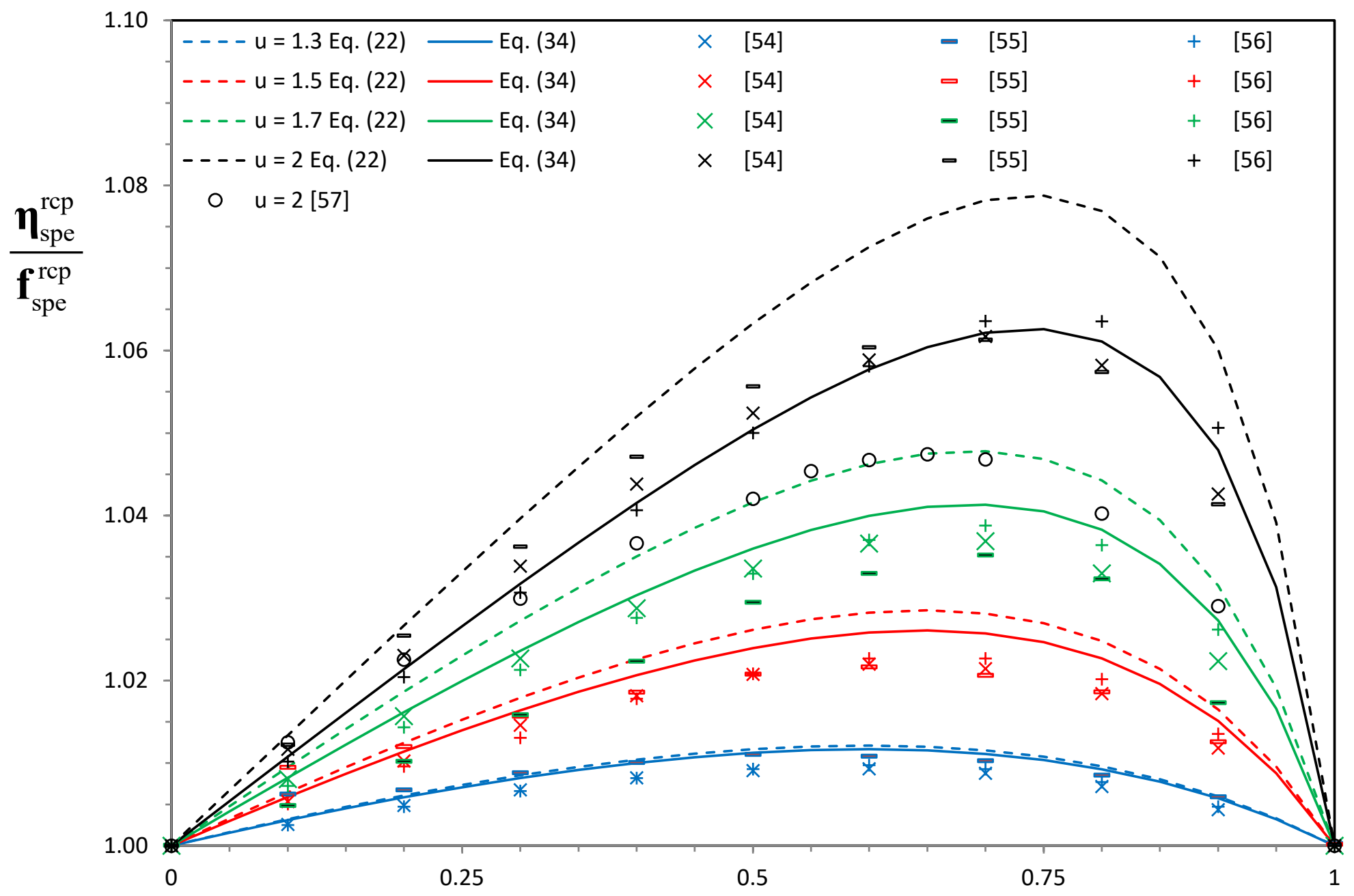

**Figure 3** Scaled packing fraction of randomly close packed binary spheres (size ratios u = 1.3, 1.5, 1.7, and 2), $\eta_{spe}^{rcp}$ (u, $c_L$)/$f_{spe}^{rcp}$, as a function of the large sphere volume fraction. Graph contains Eq. (23), with either Eq. (22) or Eq. (34) substituted, and computer-generated data (listed in Tables 1 and 3 (l/d = 0)).

A comparison with the binary RCP of spheres for larger u is offered by Desmond and Weeks [25]. They numerically generated polydisperse packings of spheres with different diameter-size distributions: binary, linear, Gaussian, and lognormal. The packing fractions of these different polydisperse packings were captured using a single equation containing the polydispersity (“δ”) and skewness (“S”). For the binary distribution, polydispersity and skewness are functions of the particle size ratio (“η”) and composition (“ρ”); see “Table I” [25]. In Table 2, the packing results of three of the generated binary packings are summarized [57].

| **“δ” [25]** | **“S” [25]** | **u [57]** | **$X_L$ [57]** | **$\eta_{spe}^{rcp}$ [57]** | **$\eta_{spe}^{rcp}$ (Eq. (34))** |
|---|---|---|---|---|---|
| 0.4 | 1.5 | 2.27 | 0.20 | 0.683 | 0.685 |
| 0.4 | 3.0 | 2.63 | 0.08 | 0.697 | 0.687 |
| 0.4 | - 0.5 | 2.70 | 0.62 | 0.653 | 0.668 |

**Table 2** Computer simulation results of RCP fraction of binary spheres [25, 57] and binary packing fraction using Eqs. (21) and (34), and $f_{spe}^{rcp}$ = 0.634.

In this table, the binary packing fractions are included, according to Eqs. (21) and (34), which are based on the number fraction $X_L$ and using $f_{spe}^{rcp}$ = 0.634 [25]. For u = 2.27, agreement is still very good; however, for larger u (2.63 and 2.70), the limitations of these equations become apparent.

### 4.2 Packings of spherocylinders

In this subsection, Eq. (23) is applied to the packing of binary spherocylinders with various aspect ratios, l/d. In the previous subsection, it was seen that Eq. (34) provides best agreement and that Eq. (22) appears to overestimate the bidisperse packing fraction for u = 2. This is no surprise: in Figure 2, it can also be seen that both contraction functions v(u) also diverged towards u = 2. Accordingly, henceforth, only Eq. (34) is used as the contraction function in Eq. (23).

In [58], dense binary packings of spherocylinders were numerically simulated for a size ratio u = 2 for aspect ratios (denoted as "w") l/d = 0 (spheres), 0.1, 0.35, 1, 1.5, and 2. In Table 3, the packing fractions, read off from "Fig. 10a" [58], versus the large volume fraction $c_L$ are summarized.

| $c_L$ | d/l = 0 | d/l = 0.1 | d/l = 0.35 | d/l = 1 | d/l = 1.5 | d/l = 2 |
|---|---|---|---|---|---|---|
| 0.0 | 0.645 | 0.672 | 0.686 | 0.659 | 0.643 | 0.629 |
| 0.1 | 0.653 | 0.678 | 0.691 | 0.667 | 0.651 | 0.638 |
| 0.2 | 0.660 | 0.683 | 0.696 | 0.673 | 0.658 | 0.644 |
| 0.3 | 0.664 | 0.686 | 0.700 | 0.680 | 0.664 | 0.652 |
| 0.4 | 0.669 | 0.690 | 0.702 | 0.684 | 0.668 | 0.656 |
| 0.5 | 0.672 | 0.693 | 0.705 | 0.687 | 0.671 | 0.659 |
| 0.55 | 0.674 | 0.694 | 0.706 | 0.689 | 0.673 | 0.659 |
| 0.6 | 0.675 | 0.695 | 0.706 | 0.689 | 0.673 | 0.658 |
| 0.65 | 0.676 | 0.693 | 0.704 | 0.688 | 0.673 | 0.658 |
| 0.7 | 0.675 | 0.692 | 0.704 | 0.687 | 0.672 | 0.657 |
| 0.8 | 0.671 | 0.687 | 0.700 | 0.681 | 0.665 | 0.649 |
| 0.9 | 0.664 | 0.680 | 0.694 | 0.672 | 0.657 | 0.635 |
| 1.0 | 0.645 | 0.672 | 0.686 | 0.659 | 0.643 | 0.629 |

**Table 3** Binary packing fraction $\eta_{spc}^{rcp}(u, c_L)$ following from computer simulation of RCP of spherocylinders with six different aspect ratios l/d and size ratio u = 2, derived from "Fig. 10a" of [58].

Their obtained $f_{spe}^{rcp}$ of the RCP of monodisperse spheres, 0.645 (Table 3, d/l = 0, $c_L$ = 0 or $c_L$ = 1), is the largest among all the computed packing results for this packing assembly (Tables 1 and 2). In Figure 3, the $\eta_{spe}^{rcp}(u, c_L)$ values of Table 3 pertaining to l/d = 0 (spheres) are included, divided by $f_{spe}^{rcp}$. It can also be seen that, from $c_L$ = 0 to $c_L$ = 0.4, the generated packing fractions are in line with all other packing data: the computer-generated packing fractions [54-56], and Eqs. (23) and (34), respectively. Beyond $c_L$ = 0.4, the protocol used in [58] starts to underpredict all other binary packing fractions, which are all consistent with each other. To quantify this anomaly, in Table 4, the ratio of $\eta_{spe}^{rcp}$(u, $c_L$) from Eqs. (23) and (34) and $\eta_{spe}^{rcp}$(u, $c_L$) from Table 3 (l/d = 0) are listed.

Indeed, the ratio is close to unity up to $c_L$ = 0.4, and then the discrepancy rapidly increases with larger $c_L$, reaching a maximum relative difference that is close to 2% (1.019), reflecting the drop in packing fraction featured in [58] that can be clearly seen in Figure 3. Towards $c_L$ = 1, the ratio tends to unity again.

| $c_L$ | A |
|---|---|
| 0.0 | 1.000 |
| 0.1 | 0.988 |
| 0.2 | 0.988 |
| 0.3 | 1.001 |
| 0.4 | 1.004 |
| 0.5 | 1.007 |
| 0.55 | 1.008 |
| 0.6 | 1.009 |
| 0.65 | 1.011 |
| 0.7 | 1.014 |
| 0.8 | 1.019 |
| 0.9 | 1.018 |
| 1.0 | 1.000 |

**Table 4** RCP packing fraction $\eta_{spe}^{rcp}$(u, $c_L$) following from Eqs. (23) and (34) divided by packing fraction $\eta_{spe}^{rcp}$(u, $c_L$) from spheres as simulated in [58], taken from Table 3 (l/d = 0, $f_{spe}^{rcp}$ = 0.645).

In Figure 4a, the binary packing fractions, as presented in [58], are set out for spherocylinders with l/d = 0.1, 035, 1, 1.5, and 2 (Table 3), as well as Eqs. (23) and (34). As expected, $f_{spc}^{rcp}$ depends on l/d. Furthermore, as with spheres (Figure 3), agreement between the simulations and model (Eqs. (23) and (34)) is very good for $c_L$ up to approximately 0.4. For a larger $c_L$, we see the same trend as for spheres; simulated packing fractions that drop and become smaller than might be expected. Accordingly, to compensate for this apparent simulation protocol anomaly, in Figure 4b, the data from Table 3 are included again but now multiplied (calibrated) with the ratios from Table 4, which are based on experiment- and theory-based expressions (*viz.* Eqs. (23) and (34)), which were independently confirmed using three different simulation protocols (see Figure 3).

a)

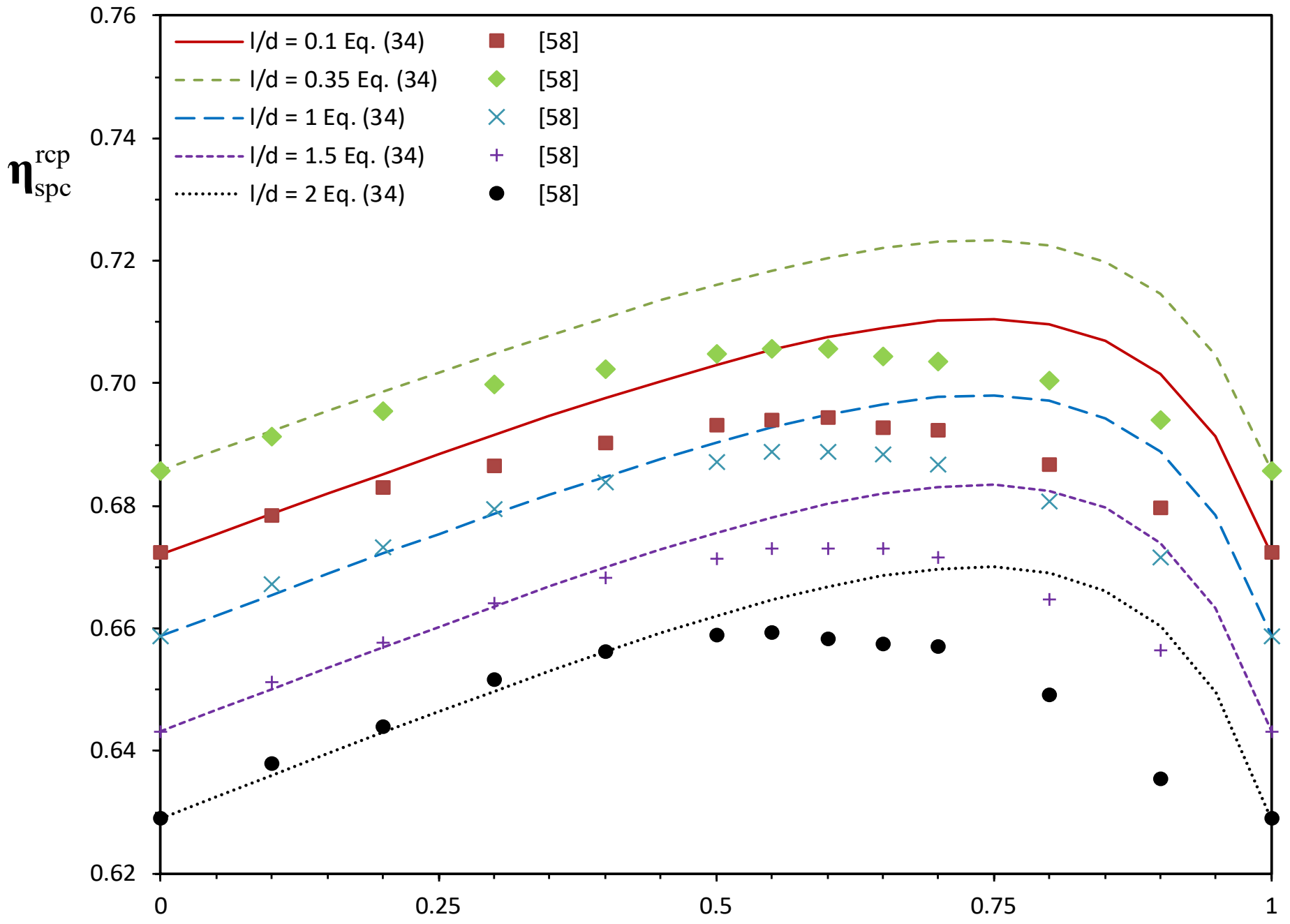

b)

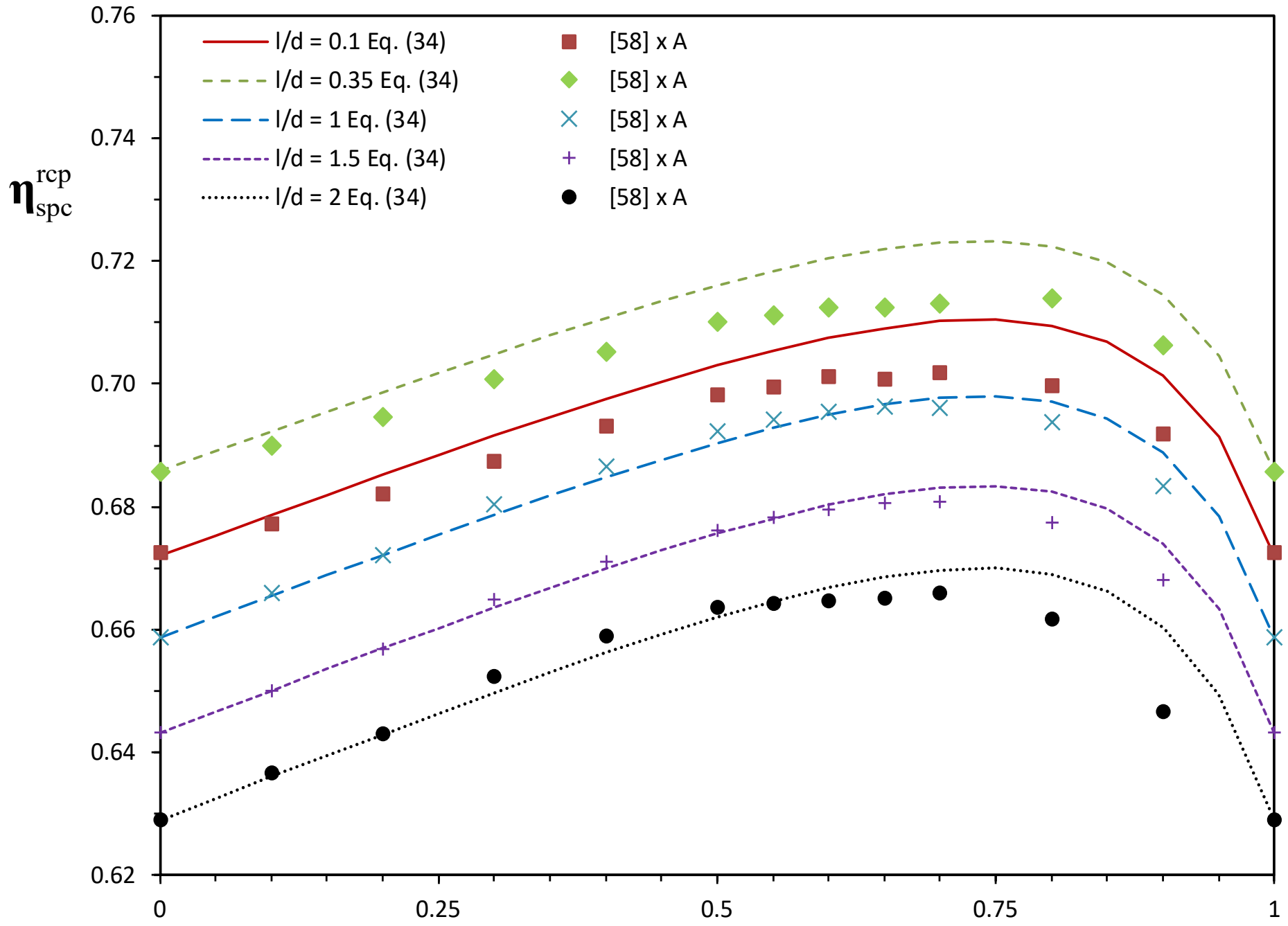

**Figure 4** Bidisperse packing fraction, $\eta_{spc}^{rcp}(u, c_L)$, of randomly close packed binary spherocylinders with size ratio u = 2 and aspect ratios l/d = 0.1, 0.35, 1, 1.5, and 2, as a function of large sphere volume fraction $c_L$. a) Eqs. (23) and (34), and computer-generated data extracted from "Fig. 10" [58] (listed in Table 3). b) Eqs. (23) and (34), and the same computer-generated data from [58] (listed in Table 3), multiplied by sphere-based adjustment A as listed in Table 4.

Figure 4b reveals that, when the data from [58] for spherocylinders with l/d ≠ 0 is adjusted with the ratio obtained from sphere (l/d = 0) packing analysis, the agreement between the simulated packing and model becomes better, and very good for l/d = 1. Graphically, the difference between the model line and some of the data points may seem large; it is of the order of 1% or so.

### 4.3 Random loose packing of spheres

The equations based on Onsager's theory revealed that the fitting coefficient C from Mangelsdorf and Washington (Eq. (29)) includes the factor $(1 - f_{spe}^{rcp})$. The Mangelsdorf and Washington expression proved to be very accurate in the entire composition range, $0 \leq c_L \leq 1$, and size ratios u up to 2 or so for a binary RCP of spheres (see Figure 3). This expression was therefore generalized to packings with different monosized packing fractions (Eqs. (23) and (34)), such as spherocylinders, and was validated in the previous subsection. Although the agreement was good (see Figure 4), the monodisperse packing fractions $f_{spc}^{rcp}$ of all considered spherocylinders did not differ significantly from those of spheres (see Table 3).

| $c_L$ | u = 1.3 | u = 1.5 | u = 1.7 | u = 2 |
|---|---|---|---|---|
| 0.0 | 0.5359 | 0.5359 | 0.5359 | 0.5359 |
| 0.1 | 0.5376 | 0.5393 | 0.5409 | 0.5429 |
| 0.2 | 0.5392 | 0.5426 | 0.5457 | 0.5500 |
| 0.3 | 0.5405 | 0.5456 | 0.5506 | 0.5571 |
| 0.4 | 0.5416 | 0.5482 | 0.5550 | 0.5641 |
| 0.5 | 0.5423 | 0.5503 | 0.5588 | 0.5707 |
| 0.6 | **0.5426** | 0.5516 | 0.5616 | 0.5764 |
| 0.7 | 0.5423 | **0.5517** | **0.5628** | **0.5803** |
| 0.8 | 0.5413 | 0.5499 | 0.5611 | 0.5802 |
| 0.9 | 0.5393 | 0.5453 | 0.5541 | 0.5711 |
| 1.0 | 0.5359 | 0.5359 | 0.5359 | 0.5359 |

**Table 5** Binary packing fraction $\eta_{spe}^{rlp}(u, c_L)$ following from computer simulation of RLP of spheres with 4 different size ratios u [56]. Maximum packing values $\eta_{spe,max}^{rlp}$ are indicated in bold.

Accordingly, in this subsection, Eqs. (23) and (34) are applied to the binary random loose packing (RLP) of spheres. This packing arrangement has a monodisperse packing fraction which is very distinct from RCP, with $f_{spe}^{rlp} \approx 0.54$ as a generally accepted value for this lower limit of random sphere packing [7, 14, 23, 29-34, 56]. The magnitude of $(1 - f)$, as a factor introduced in Eq. (23), is therefore about 0.36 and 0.46 for RCP and RLP, respectively, and this difference in the relative expression is even larger than their difference in f, *viz.* 0.64 and 0.54, respectively.

In Figure 5, the packing simulation results from the RLP of the spheres generated by [23, 56] are presented, which are listed in Table 5. In this figure, Eqs. (23) and (34) were included using $f_{spe}^{rlp}$ = 0.536, taken from Table 5.

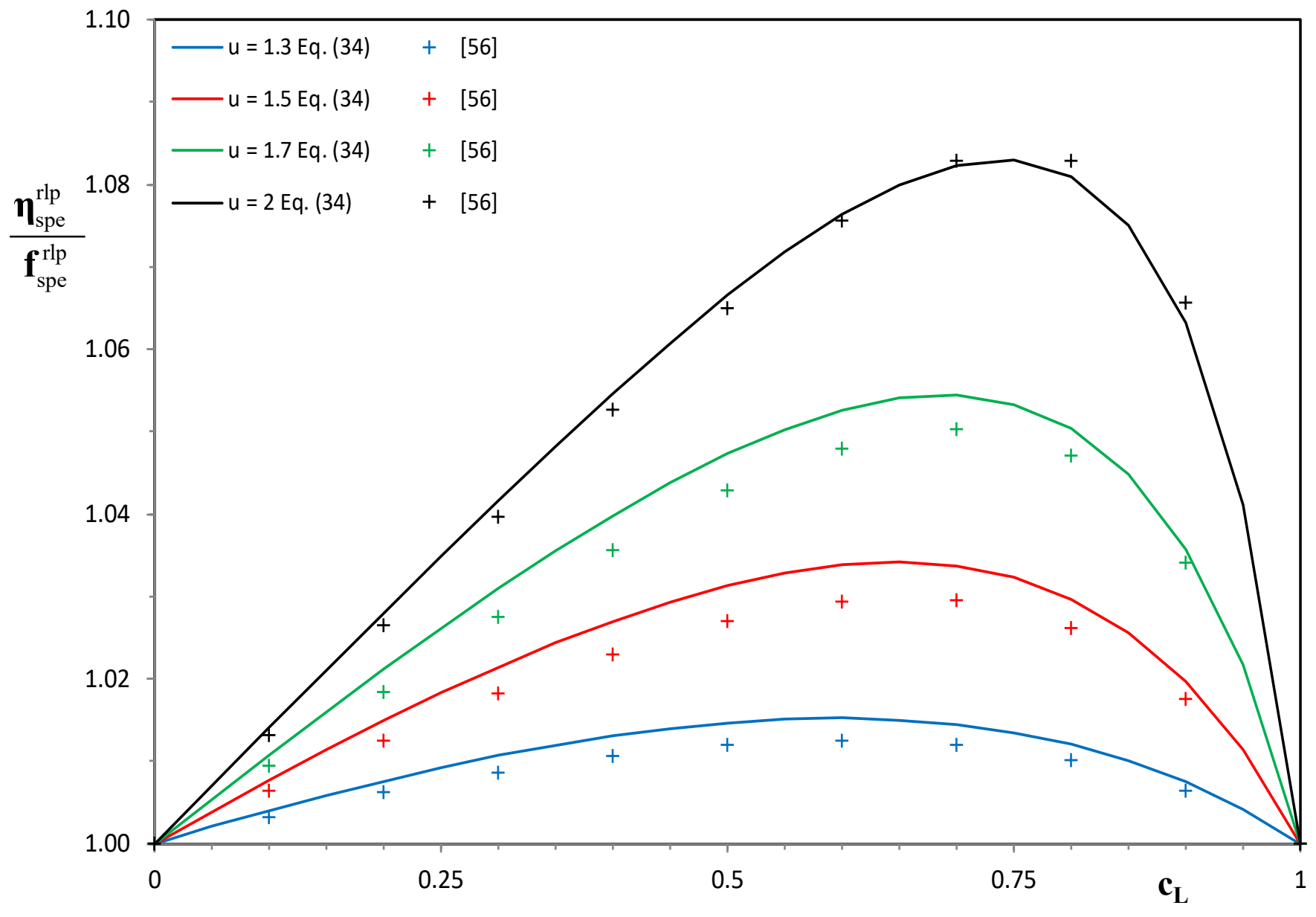

**Figure 5** Scaled packing fraction of randomly loose packed (RLP) binary spheres (size ratios u = 1.3, 1.5, 1.7, and 2), $\eta_{spe}^{rlp}(u, c_L)/f_{spe}^{rlp}$, as a function of the large sphere volume fraction. Graph contains Eqs. (23) and (34), using $f_{spe}^{rlp}$ = 0.536, and computer-generated data (listed in Table 5).

Figure 5 confirms that Eqs. (23) and (34) can accurately predict the binary RLP packing fraction of spheres in the entire compositional range $0 \leq c_L \leq 1$ and for the size ratio range $1 \leq u \leq 2$. This implies that Eqs. (23) and (34) are not only applicable to RCP, but also to RLP.

Based on the results shown in Figures 3-5 and the information provided by applying Onsager's excluded volume model, it can be tentatively concluded that Eq. (21) (or Eq. (23)), in combination with Eq. (34), is applicable to the loosest and densest packing states (RLP and RCP, respectively), and probably also for all intermediate states of compaction. It also seems applicable to spherocylinders with variable l/d, and probably to other particle types as well, although there is no rigorous proof yet for this claim. Remarkably, the present model is based solely on analytical analysis (without introducing a fitting parameter). The only parameter, monosized packing fraction f, is physically defined and is a function of only the considered particle type and mode of packing (*e.g.* loose, close). In Section 6, the monodisperse packing fraction and the effect of compaction are analyzed in more detail. In Appendix B, the insight that the factor $1 - f$ is crucial in the contraction term is translated to 2-dimensional binary disk packings and to binary hypersphere packings. In the next section, the difference in contraction functions v(u), such as provided by the Onsager excluded volume model, the Mangelsdorf and Washington model, and [37], and, accordingly Eqs. (22), (34) and (35), is further analyzed.

## 5. Analysis of models

As seen before, for u → 1, the Onsager- and Mangelsdorf- and Washington-based expressions coincide as Eqs. (22) and (34) coincide, as demonstrated in Subsection 3.2 and shown in Figure 2. On the other hand, for larger size differences (u deviating more from unity), the volume contraction functions begin to diverge. This contraction function governs the 3rd term in the denominator of Eqs. (21) and (23). The resulting difference in the predicted binary packing fraction is analyzed in more detail.

To characterize the effect of the different v(u) on the binary packing fraction $\eta(u, X_L)$, the maximum packing fraction according to both approaches, being a function of the size ratio u only, is determined. For a given u, the extremum in Eq. (21) follows from the partial derivative of $\eta(u, X_L)$, with respect to $X_L$:

$$\eta_{X_L}(u, X_L) = \frac{f((X_L - 1)^2 - X_L^2 u^3)v(u)}{\left(X_L(u^3 - 1) + 1 - (1 - f)X_L(1 - X_L)v(u)\right)^2} \quad . \tag{37}$$

Equating Eq. (37) to zero yields the number fraction $X_{L,max}$ and volume fraction $c_{L,max}$, which results in the maximum packing for a given size ratio u:

$$X_{L,max} = \frac{1}{u^{3/2} + 1} \quad ; \quad c_{L,max} = \frac{u^{3/2}}{u^{3/2} + 1} \quad , \tag{38}$$

where $c_{L,max}$ is obtained using Eq. (24), but it also follows from Eq. (23) and solving the $\eta_{c_L}(u, c_L) = 0$ equation. Note that these specific $X_L$ and $c_L$ values do not depend on v(u) or on the monosized packing fraction f. Eq. (38) also reveals that, as u > 1, the maximum packing fraction is found for $X_{L,max} < 0.5$, and $c_{L,max} > 0.5$.

Figures 3-5 indeed show that the packing fraction is maximum for the larger particle volume fraction $c_L$ exceeding parity, and that the eccentricity in the optimal packing composition increases with increasing u, as predicted by Eq. (38). It is also interesting to note that $X_{L,max} + c_{L,max} = 1$, and, hence, $X_{L,max}$ and $c_{L,max}$ are symmetrical with respect to 0.5 for all u. In Figure 6, they are both set out against u, as well as $c_{L,max}$, which follows from Tables 1 and 5 (values indicated in bold). As the packing fraction values are computed with $c_L$ intervals of 0.1, $c_{L,max}$ (and maximum packing fraction $\eta_{max}$) can only be roughly estimated. For example, in Table

1c), one can see for u = 2 that the maximum packing fraction is located between $c_L = 0.7$ and 0.8; hence, $c_{L,max} = 0.75$ is shown in Figure 6 as an approximation.

The composition corresponding to the maximum particle packing is a function of u only, and the corresponding maximum packing follows by substituting $X_{L,max}$ from Eq. (38) into Eq. (21), or $c_{L,max}$ from Eq. (38) into Eq. (23), and both yield

$$\frac{\eta_{max} - f}{\eta_{max}\,(1 - f)} = \frac{v(u)}{(u^{3/2} + 1)^2} \quad . \tag{39}$$

This scaled maximum of the binary packing fraction does not depend on the monosized packing fraction f of the considered particle and contains the factor ($\eta_{max}$ – f), which reflects the maximum packing increase due to binary polydispersity. One can also identify (1 – f), which is the factor that follows from the analysis of Onsager's equations.

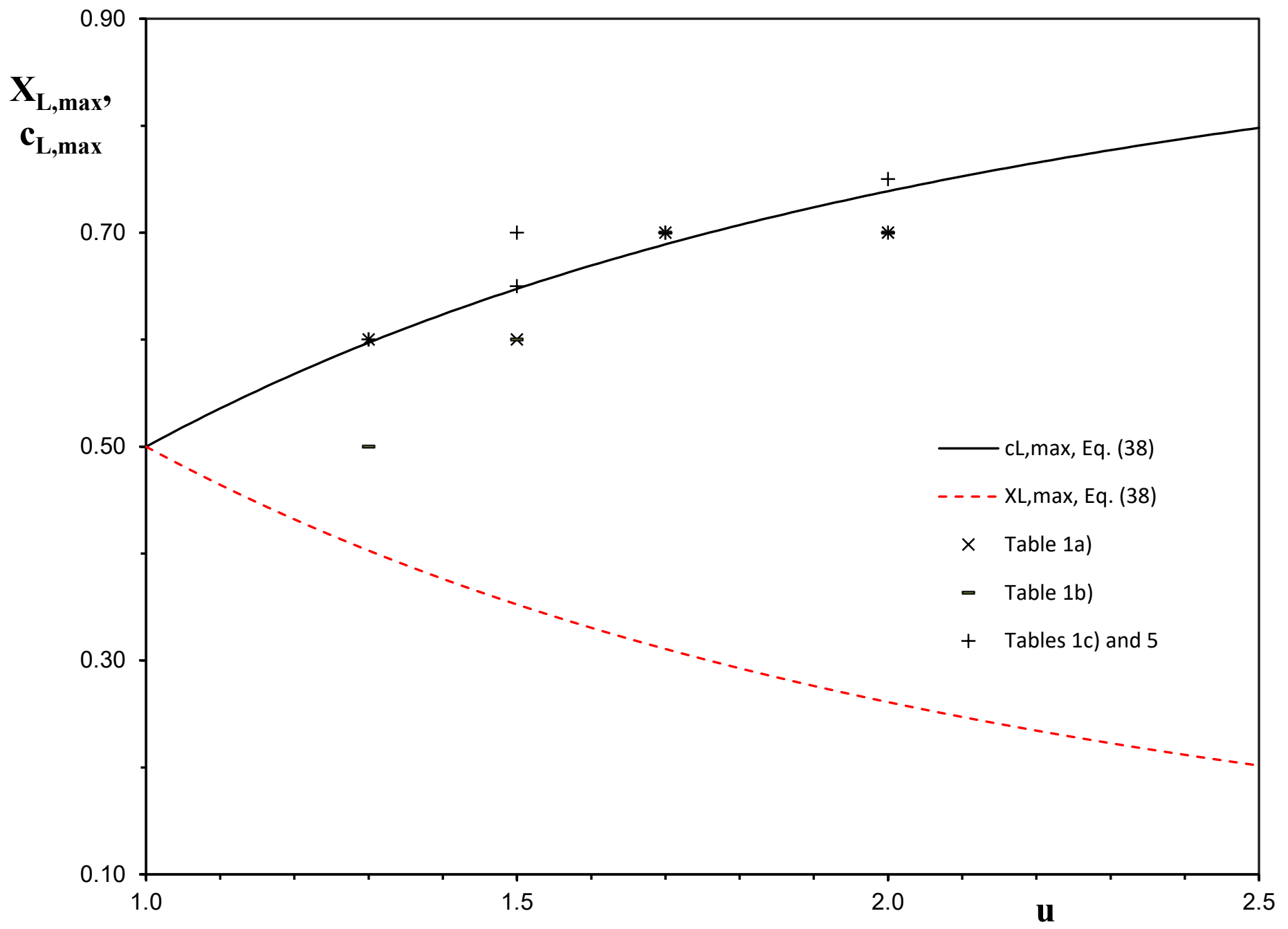

**Figure 6** Volume ($c_{L,max}$) and number ($X_{L,max}$) fraction, Eq. (38) of large spheres, which result in a maximum packing fraction for bidisperse particle mixes, as a function of size ratio u. Both functions are symmetrical with respect to the horizontal line with a value of 0.5. In this figure, $c_{L,max}$ values from Tables 1 and 5 are also included.

To illustrate the effect of v(u) given by Eqs. (22), (34), and (35), the resulting ($\eta_{max}$ – f)/$\eta_{max}$(1 – f) values are shown in Figure 7. The difference in the packing maximum predicted by the Onsager- and Mangelsdorf and Washington-based models for a given u is very minor for small (u – 1), which is not surprising, as v(u) from both Eqs. (22) and (34) tends to $2(u – 1)^2$ for $u \to 1$. However, for larger u, the difference becomes more pronounced, and at u = 2, it is approximately 25%, which is also seen in Figure 2. Figure 7 also shows the linear approximation of the contraction function (Eq. (35)) is close to second-order models based on the Onsager model and Mangelsdorf and Washington experiments, but is less accurate in the entire range of $1 \leq u \leq 2$. For small $(u - 1)$, owing to its linear dependence on $u - 1$, it overestimates the packing fraction, whereas for $u \rightarrow 2$ and larger, it underestimates the binary packing fraction.

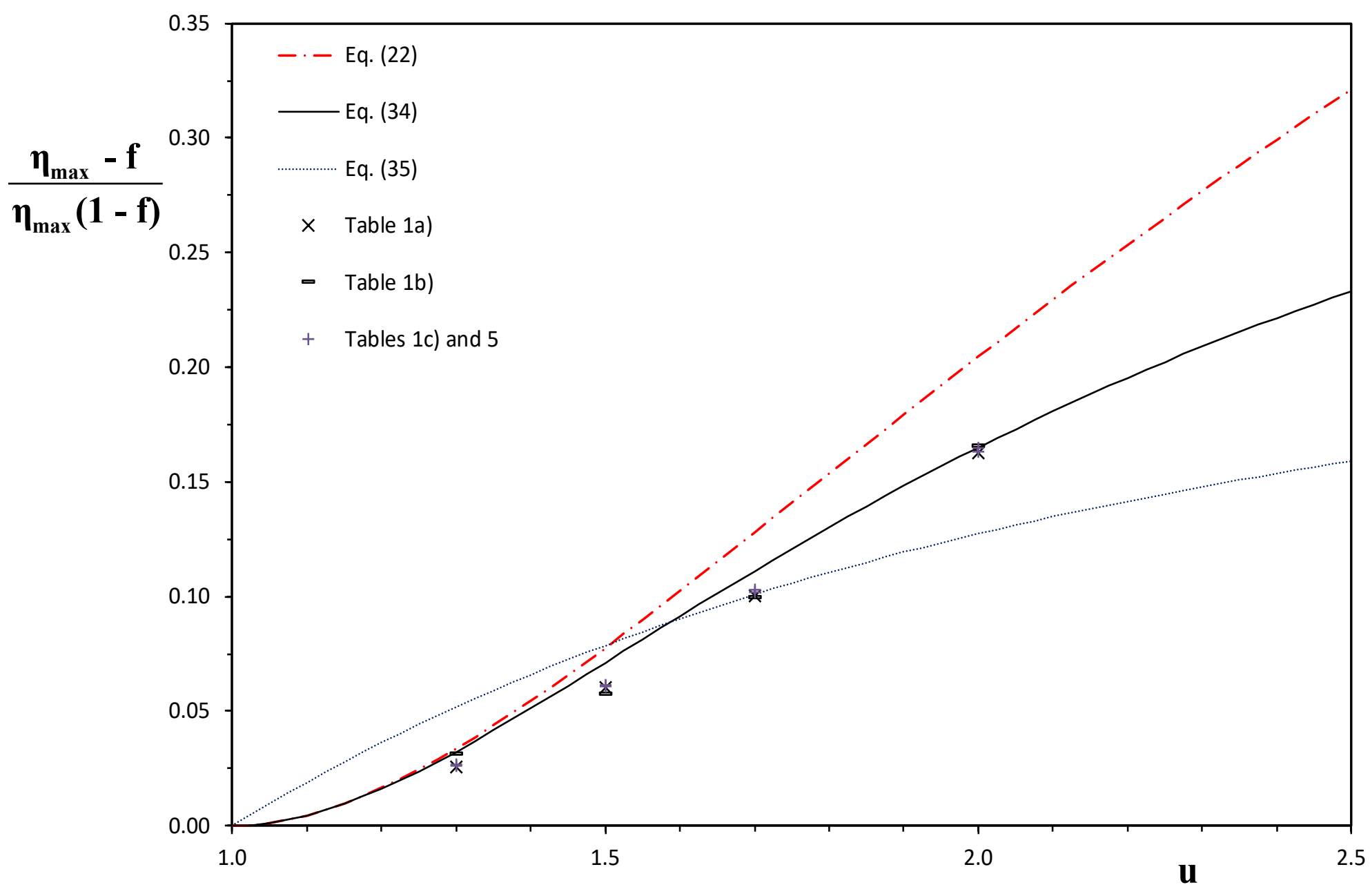

**Figure 7** Scaled highest binary packing fraction, $(\eta_{max} - f)/\eta_{max}(1 - f)$, as a function of size ratio u, u ranging from unity to two, computed with Eq. (39) and using Eqs. (22), (34), and (35); for the last $\beta = 0.20$ is employed [37, 53]. In this figure, scaled maximum packing fractions extracted from Tables 1 and 5 are also included.

Figures 3-5 showed that Eqs. (23) and (34) capture the binary packing fraction very well over the entire considered size ratio range of $1 \leq u \leq 2$. That Eq. (23) with Eq. (34) best captures the binary packing fraction is also confirmed by the scaled packing maxima that follow from the packing simulations (taken from Tables 1 and 5), which are included in Figure 7. The figure also shows that the scaled RLP packing maxima coincide with RCP values. This conclusion is not obvious, and once more emphasizes the importance of the scaling factor $(1 - f)$ that followed from comparing the Onsager- and Mangelsdorf and Washington-based models (Section 3). The agreement of the different simulation protocols illustrates that the packing fractions as such may differ (Tables 1 and 5), but that $(\eta_{max} - f)/\eta_{max}(1 - f)$ are very similar. In other words, though the computed monosized and bidisperse packing fractions differ, the maximum effect

of bidisperse polydispersity on the scaled packing fraction, governed by $(\eta_{max} - f)/\eta_{max}(1 - f)$, is almost identical.

## 6. Packing fraction of monodisperse particles

In Section 3, it was seen that the Onsager-based model predicts a packing increase by bidispersity that amounts to $2f(1 - f)X_1(1 - X_1)(u - 1)^2 + O((u - 1)^3)$, Eq. (31). The monodisperse packing fraction f depends on particle type and its densification (Appendix A, for instance, addresses the monodisperse packing fraction of (sphero)cylinders with different l/d, that is, different shapes). Monodisperse particles can be packed between the loosest and closest states of densification, the packing fraction being $f_{\square}^{\mathrm{rlp}} \le f \le f_{\square}^{\mathrm{rcp}}$. This implies that the factor 2f(1 - f) depends on the packing configuration. In this section, the range of f is investigated, while a map of known $f_{\square}^{\mathrm{rlp}}$ and $f_{\square}^{\mathrm{rcp}}$ combinations for a number of particle type is introduced.

### 6.1 Packing fraction range of a particle type

The packing fraction of particles may range from their loosest possible state of packing (RLP) to their densest state of packing (RCP). Figure 8 depicts combinations of ($f_{\square}^{\mathrm{rlp}}$, $f_{\square}^{\mathrm{rcp}}$) for a number of particle types, taken from the literature. For spheres, the point ($f_{\mathrm{spe}}^{\mathrm{rlp}} = 0.54$, $f_{\mathrm{spe}}^{\mathrm{rcp}} = 0.64$) is included. In Table 6, the RLP and RCP packing fractions of the five platonic solids, taken from [59], are listed and are included in Figure 8 as well. These authors of [62] experimentally determined the RLP and RCP packing fractions of the monodisperse five platonic solids: tetrahedron, cube, octahedron, dodecahedron, and icosahedron, all having slightly rounded edges.

| **Shape** | $f_{\square}^{\mathrm{rlp}}$ | $f_{\square}^{\mathrm{rcp}}$ |
|---|---|---|
| Tetrahedron | 0.36 | 0.49 |
| Cube | 0.33 | 0.46 |
| Octahedron | 0.36 | 0.48 |
| Dodecahedron | 0.37 | 0.49 |
| Icosahedron | 0.41 | 0.50 |

**Table 6** Random loose and close packing fractions of five platonic solids as measured by [59].

Furthermore, for a number of disks (l/d < 1), equilateral cylinders (l/d = 1), cylinders (l/d > 1), and particles shaped otherwise, the RLP and RCP void fractions (= 1 - f) were measured by [60], depicted in “Fig. 2a” and “Fig. 2b”, respectively, versus the Wadell sphericity. The disks’ l/d is taken from “Table 1”, and the cylinders’ l/d from the main text [60]. From these l/d, the Wadell sphericity is computed (Table 7), which matches excellently with the Wadell sphericity from “Fig. 2” [60]. Hence, for each particle, $f_{\square}^{rlp}$ and $f_{\square}^{rcp}$ can be derived from their “Fig. 2a” and “Fig. 2b” [60], respectively, which are summarized in Table 7, and they are included in Figure 8 as well.

| **Cylinders** | | | | **Disks** | | | | **Particles shaped otherwise** | | |
|---|---|---|---|---|---|---|---|---|---|---|
| **Ψ** | **l/d** | $\mathbf{f_{cyl}^{rlp}}$ | $\mathbf{f_{cyl}^{rcp}}$ | **Ψ** | **l/d** | $\mathbf{f_{cyl}^{rlp}}$ | $\mathbf{f_{cyl}^{rcp}}$ | **Ψ** | $\mathbf{f_{\square}^{rlp}}$ | $\mathbf{f_{\square}^{rcp}}$ |
| 0.874 | 1 | 0.590 | 0.702 | 0.594 | 0.1656 | 0.563 | 0.671 | 0.862 | 0.613 | 0.644 |
| 0.734 | 4 | 0.534 | 0.636 | 0.520 | 0.1229 | 0.553 | 0.635 | 0.821 | 0.598 | 0.671 |
| 0.677 | 5.6 | 0.481 | 0.610 | 0.352 | 0.0579 | 0.418 | 0.558 | 0.801 | 0.604 | 0.658 |
| 0.617 | 8 | 0.463 | 0.582 | 0.200 | 0.0225 | 0.361 | 0.469 | 0.793 | 0.588 | 0.671 |
| 0.561 | 11.2 | 0.428 | 0.515 | 0.169 | 0.0172 | 0.307 | 0.433 | 0.769 | 0.570 | 0.624 |
| 0.504 | 16 | 0.310 | 0.436 | 0.133 | 0.0119 | 0.285 | 0.403 | 0.737 | 0.566 | 0.645 |
| 0.455 | 22.4 | 0.269 | 0.345 | 0.094 | 0.0094 | 0.270 | 0.384 | 0.723 | 0.564 | 0.645 |
| 0.406 | 32 | 0.171 | 0.234 | | | | | 0.684 | 0.567 | 0.655 |
| 0.365 | 44.8 | 0.101 | 0.142 | | | | | 0.568 | 0.504 | 0.596 |
| 0.325 | 64 | 0.063 | 0.097 | | | | | 0.554 | 0.574 | 0.651 |
| | | | | | | | | 0.450 | 0.307 | 0.403 |
| | | | | | | | | 0.423 | 0.430 | 0.523 |
| | | | | | | | | 0.413 | 0.212 | 0.303 |
| | | | | | | | | 0.285 | 0.322 | 0.425 |

**Table 7** Measured RCP and RLP packing fraction values for different particle types, the values of ratio l/d ($\alpha^{-1}$), and Wadell sphericity Ψ values, derived from “Fig. 2” of [60]. Disks’ l/d is taken from “Table 1”, and cylinders’ l/d, from the main text [60]. For cylinders and disks, $\Psi = 2\,(3/2)^{2/3}\,\alpha^{1/3}/(2 + \alpha)$.

One can see in Figure 8 that for all particles, $f_{\square}^{rcp} > f_{\square}^{rlp}$, as would be expected, so all ($f_{\square}^{rlp}$, $f_{\square}^{rcp}$) are located above the diagonal line $f_{\square}^{rcp} = f_{\square}^{rlp}$ (parity line in Figure 8). From basic principles, it is required that $f_{\square}^{rlp}$ and $f_{\square}^{rcp}$ coincide when $f_{\square}^{rcp}$ and $f_{\square}^{rlp}$ approach zero and unity, respectively, since $0 \le f_{\square}^{rlp} < f_{\square}^{rcp} \le 1$. One can observe that the difference between $f_{\square}^{rcp}$ and $f_{\square}^{rcp}$ indeed diminishes when they tend to zero and unity. It also appears that all combinations of $f_{\square}^{rlp}$ and $f_{\square}^{rcp}$ are located below a curve that obeys ($f_{\square}^{rlp} = 0$, $f_{\square}^{rcp} = 0$) and ($f_{\square}^{rlp} = 1$, $f_{\square}^{rcp} = 1$). Hence, it is plausible to propose as *ansatz* an upper bound on the random close packing value, depending on the random loose packing value, as

$$f_{\square}^{rcp} = f_{\square}^{rlp} + B f_{\square}^{rlp}(1 - f_{\square}^{rlp}) \quad , \tag{40}$$

with $B > 0$, which is a parabola superimposed on the parity line $f_{\square}^{rcp} = f_{\square}^{rlp}$. The equation furthermore fulfills the conditions $f_{\square}^{rcp} > f_{\square}^{rlp}$, $f_{\square}^{rlp} = f_{\square}^{rcp} = 0$, and $f_{\square}^{rlp} = f_{\square}^{rcp} = 1$, and hence has the good characteristics of an upper threshold. The inverse relation $f_{\square}^{rlp}$ for a given $f_{\square}^{rcp}$ follows from solving Eq. (40), yielding

$$f_{\square}^{rlp} = \frac{1+B-\sqrt{(B+1)^2 - 4Bf_{\square}^{rcp}}}{2B} \quad . \tag{41}$$

Also, this expression obviously fulfills the conditions $f_{\square}^{rlp} \leq f_{\square}^{rcp}$, $f_{\square}^{rcp} = f_{\square}^{rlp} = 0$, and $f_{\square}^{rcp} = f_{\square}^{rlp} = 1$. It appears that all ($f_{\square}^{rlp}$, $f_{\square}^{rcp}$) values are below Eq. (40) for B ≈ 0.6, which is a putative maximum (Figure 8).

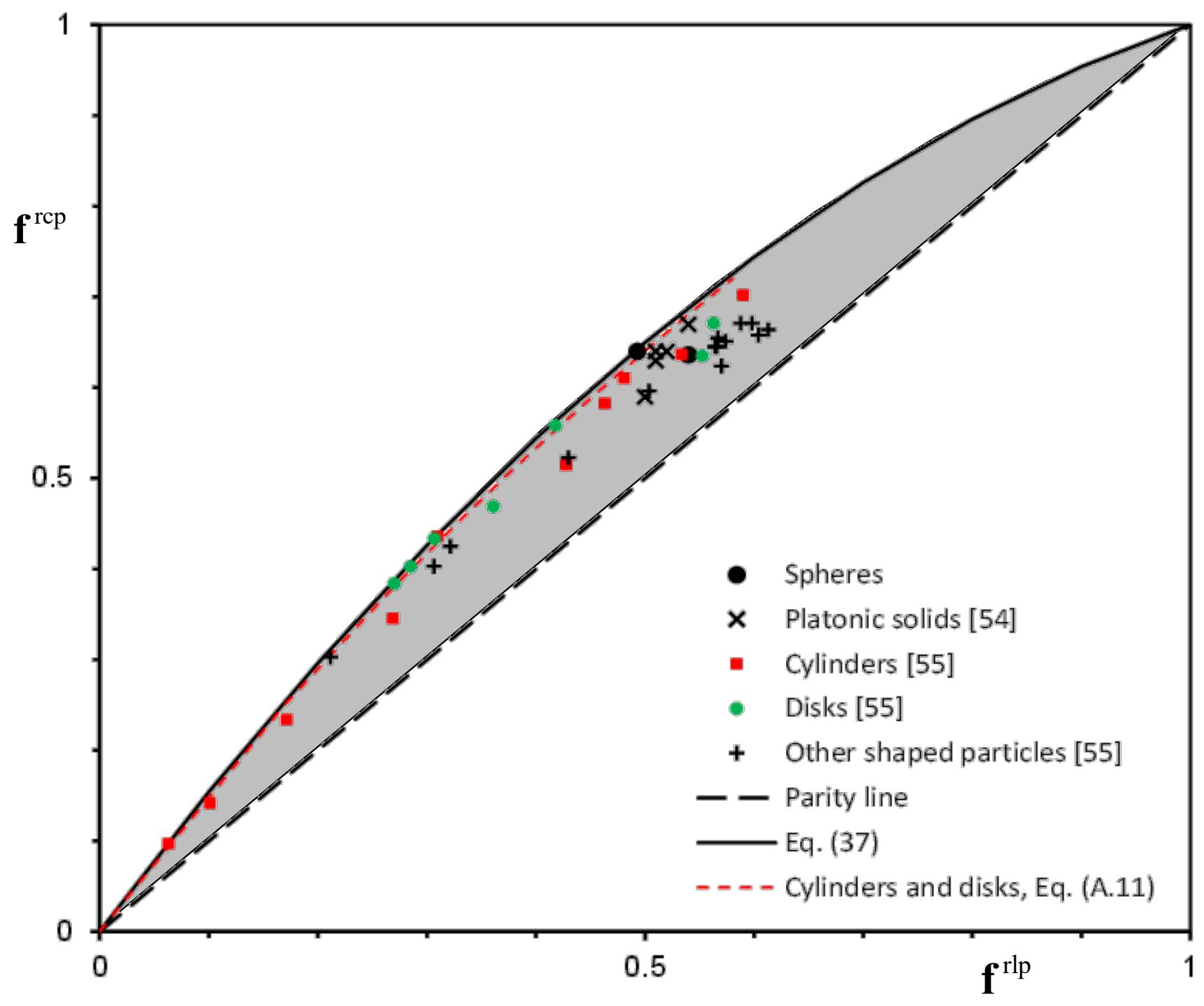

**Figure 8** Packing fraction map of monosized particles showing close packing fraction $f_{\square}^{rcp}$ as a function of their loose packing fraction $f_{\square}^{rlp}$ (shaded area). Curve pertaining to maximum packing fraction, $f_{\square}^{rcp}$, is shown using Eq. (40) with B = 0.6, as well as measured combinations of $f_{\square}^{rlp}$ and $f_{\square}^{rcp}$ values for a number of particle types, taken from Tables 6 and 7. For spheres taken as ($f_{spe}^{rlp}$, $f_{spe}^{rcp}$), values (0.54, 0.64) and (0.49, 0.64) are included (text). For fitted cylinder packing fractions ($0 \leq l/d < \infty$), Eq. (A.11) for RLP and RCP is included as well.

The factor B characterizes the curvature of the upper threshold. Eq. (40) with B = 0.6 is included in Figure 8, and it can be seen that, indeed, almost all combinations of ($f_{\square}^{rlp}$, $f_{\square}^{rcp}$) are located in the area (shaded) enclosed by line $f_{\square}^{rcp} = f_{\square}^{rlp}$ and Eq. (40). Many points are located on the curve, despite the fact that the packing fraction of different particle types, but with identical Ψ, may differ (see "Fig. 2" [60] and Table 7). Many $f_{\square}^{rcp}$ and $f_{\square}^{rlp}$ seem to obey Eq. (40).

Many combinations of ($f_{\square}^{rlp}$, $f_{\square}^{rcp}$) are located within the shaded area. It is conceivable, though still rather speculative, to believe that, for these particles, in order to have ($f_{\square}^{rlp}$, $f_{\square}^{rcp}$) located on the threshold line, either the true RLP or RCP packing configurations of such particles were not measured or computed, or neither was measured or computed. There is no proof for this conjecture, but if the actual $f_{\square}^{rcp}$ of such particles were higher, the data point would move upward, and if the actual $f_{\square}^{rlp}$ were lower, it would move to the left. In both cases, the threshold curve is then approached.

In this respect, a nice example is ($f_{spe}^{rlp}$, $f_{spe}^{rcp}$) of spheres. The depicted combination of (0.54, 0.64) is located within the area and not on the proposed boundary (Eq. (40)). The RCP fraction of spheres is undisputedly close to this value of 0.64, but the value for the RLP is less agreed upon. To be located on curve Eq. (40), the required $f_{\square}^{rlp}$ has to follow from Eq. (41) by substituting $f_{spe}^{rcp}$= 0.64 and B = 0.6, yielding $f_{spe}^{rlp}$ = 0.49. Interestingly, in [28], an RLP packing fraction lower than 0.54 was indeed proposed, namely, the lowest packing fraction of 0.49 for strictly jammed frictionless spheres, a value that would put ($f_{spe}^{rlp}$, $f_{spe}^{rcp}$) on the boundary curve. This second spheres point ($f_{spe}^{rlp}$, $f_{spe}^{rcp}$) = (0.49, 0.64) is therefore added to Figure 8 as well.

In Figure 8, the RCP and RLP packing fractions of cylinders for 0 ≤ l/d ≤ 180, based on Eq. (A.11), are included as well. For these l/d, the packing fraction ($f_{cyl}^{rlp}$, $f_{cyl}^{rcp}$) range from (0, 0) to maxima (0.58, 0.72) (see Appendix A). The former combination is attained for l/d = 0 ($\alpha^{-1}$), and the maximum, at l/d = 0.384 ($\alpha_{max}$ = 2.6). Figure 8 reveals that the combination of fitted RLP and RCP packing fractions follows threshold curve Eq. (40) very closely for all l/d, whereby each l/d (or α) represents a distinct particle type.

**6.2 Effect of packing configuration on bidisperse packing**

Based on putative threshold Eq. (40), it is possible to assess the range of f(1 - f), which is a factor in the bidisperse packing increase (e.g. Eqs. (31) and (36)) for f ranging from $f_{\square}^{rlp}$ to $f_{\square}^{rcp}$, depending on the applicable packing configuration. Using Eq. (40), it follows that

$$f_{\square}^{rcp}(1 - f_{\square}^{rcp}) = f_{\square}^{rlp}(1 - f_{\square}^{rlp})(1 - Bf_{\square}^{rlp})(1 + B - Bf_{\square}^{rlp}) \quad , \qquad (42)$$

As $f_{\square}^{rlp} \leq f \leq f_{\square}^{rcp}$, now the effect of densification on the range of f(1 - f) values can also be quantified. Eq. (42) reveals that the RCP/RLP ratio of the product f(1 - f) is $(1 - Bf^{rlp})(1 + B - Bf^{rlp})$. By differentiation, it readily follows that this function is decreasing from $f^{rlp} = 0$ to $f^{rlp} = 1$, at $f^{rlp} = 0$ it is 1.6, and at $f^{rlp} = 1$, $(1 - Bf^{rlp})(1 + B - Bf^{rlp}) = 0.4$. Hence, for particles with larger $f^{rlp}$ (and larger $f^{rcp}$), a denser configuration will result in a higher f(1 - f). On the other hand, for particles with small $f^{rlp}$, *e.g.* (sphero)cylinders with high aspect ratio l/d (Appendix A), loose packings will have a higher f(1 - f) than their denser configuration. In other words, the effect of bidispersity on the packing increase will then be more pronounced.

Now, the question arises as to whether the packing fraction of a bidisperse close packing always exceeds that of the loose bidisperse particle packing, both having the same composition and size ratio. In that case

$$f^{rcp} + 2f^{rcp}(1 - f^{rcp})X_L(1 - X_L)(u - 1)^2 > f^{rlp} + 2f^{rlp}(1 - f^{rlp})X_L(1 - X_L)(u - 1)^2 \quad . \qquad (43)$$

Substituting Eqs. (40) and (42) here yields the requirement

$$B + 2(1 - Bf^{rlp})(1 + B - Bf^{rlp})X_L(1 - X_L)(u - 1)^2 > 2X_L(1 - X_L)(u - 1)^2 \quad . \qquad (44)$$

As $(u - 1)^2 \leq 1$, $X_L(1 - X_L) \leq ¼$, and $B = 0.6$, Eq. (44), and consequently Eq. (43) as well, is always fulfilled. Hence, for each particle type and $u \leq 2$, bidispersity cannot lead to its RLP packing fraction exceeding its RCP packing fraction.

## 7. Concluding remarks

In this study, the packing fraction of polydisperse packings was addressed, in particular, the packing of binary assemblies with a limited size ratio u (maximum 2 to 2.5 or so), obtained by combining two uniformly shaped (similar) particles of different sizes. Here, the packing of random bidisperse particles with small size differences is studied in detail by revisiting Onsager's theory, which is based on excluded volume and was originally developed to model the liquid-to-nematic (I-N) phase transition of hard rod-like (spherocylinders and cylinders) particles [1].

The Onsager expressions for the excluded volume have so far only been used to study the packing of identical particles, although his theory allows for two (sphero)cylinders with different diameters and lengths. In this study, it is applied to the packing of bidisperse similar (sphero)cylinders, *i.e.* with the same aspect (length/diameter) ratio $\alpha$ (= d/l).

Using a statistical approach, in Section 2, binary Onsager expressions are used to derive a closed-form expression for the excluded volume and, subsequently, the binary packing fraction of the aforesaid particles. It appears that the expressions for both spherocylinders and cylinders are identical (Eqs. (21) and (23)), irrespective of the aspect ratio. From a Taylor expansion, it follows that the bidisperse variation from the monosized packing fraction amounts to $2f(1 - f)X_1(1 - X_1)(u - 1)^2 + O((u - 1)^3)$; hence, it is a function of the concentration $X_1$ and proportional to the square of the relative size difference u – 1. In Section 3, a bidisperse packing fraction expression is derived based on a semi-empirical model of the RCP of spheres [2]. It is shown that, for $u \rightarrow 1$, this semi-empirical expression coincides with the expression obtained from using Onsager's excluded volume model, which includes spheres. It furthermore appears that the fitting parameter C [2] is uniquely related to (1 – f).

To validate and compare the obtained expressions for the packing increase, in Section 4, they are first compared with numerical simulations of the binary RCP of spheres [54-56] over the entire compositional range (concentration from zero to unity) and for size ratios u up to well over two [57]. For a size ratio u value towards 2, Eq. (34), following [2], is more accurate than Eq. (22) following Onsager's model of excluded volume [1], resulting in Eqs. (21) and (23) in combination with Eq. (34) being a general bidisperse packing fraction applicable to u up to 2–2.5. Subsequently, a comparison with RCP for spherocylinders with l/d ranging from 0.1 to 2, and u = 2, again yields good agreement. Finally, the general packing expression is applied to and compared with the computationally generated RLP of spheres [56], for which the factor (1

– f) is 25% larger than for RCP, and excellent agreement is found in the entire compositional range and for size ratios u up to 2.

In Section 5, the different approaches to the binary packing fraction, *viz.* [1], [2] and [37], are mutually compared by computing their scaled maximum packing fraction $(\eta_{max} - f)/\eta_{max}(1 - f)$, which no longer depends on monodisperse packing fraction f but on the contraction function v(u) and size ratio u only (Eq. (39)). In Figure 7, the scaled maximum packing fractions of these models, which have different contraction functions v(u), are set out, as are the corresponding results of computer simulations (both RCP and RLP of spheres), again confirming that the combination of Eq. (21) (or Eq. (23)) and Eq. (34) yields accurate results.

In Section 6, the magnitude of the factor f(1 - f) of a particle type, governing the increase in its packing fraction by bidispersity, is assessed by considering the packing fraction range, which may vary from $f_{\square}^{rlp}$ to $f_{\square}^{rcp}$. First, a packing fraction map is introduced (Figure 8) that covers the area of known possible combinations of $f_{\square}^{rlp}$ and $f_{\square}^{rcp}$ from a number of particle types. An explicit threshold (Eq. (40)) is proposed, and subsequently the effect of the packing configuration on the factor f(1 - f) is clarified. Eq. (37), with B = 0.6, appears to be an upper threshold of ($f_{\square}^{rlp}$, $f_{\square}^{rcp}$) combinations of reported monodisperse loose and close packing data of many different particle types, and many ($f_{\square}^{rlp}$, $f_{\square}^{rcp}$) combinations are located at this threshold (Figure 8).

Appendix A contains a literature review of monodisperse (sphero)cylinder packing data on RLP and RCP. It appears that the monodisperse packing expressions following from Onsager's excluded volume, Eqs. (6) and (7), are in qualitative agreement only. This is not a surprise, as Onsager's theory was not developed for assessing the packing fraction of monosized RCP or RLP packings. Notwithstanding this conclusion for the monodisperse situation, this paper reveals that the excluded volume approach yields a *quantitative correct expression for the bidisperse case*. Apparently, the concept of excluded volume is able to capture the volume contraction of a packing assembly by studying the excluded volume of all particle pairs, which are randomly oriented and randomly composed (Figure 1). This volume reduction of excluded volumes and, hence, the entire assembly, exceeds the particle volume reduction involved by the introduction of smaller particles, resulting in the observed packing fraction increase by the binary polydispersity. Or, alternatively formulated, the introduction of larger particles increases the excluded volume less than that of the larger particle volume. Elaborating further on this conclusion, in Appendix B, expressions are proposed for binary hyperspheres following the excluded volume theory of Onsager, accompanied by the introduction of a Mangelsdorf and Washington-type contraction function.

As stated in the Introduction, for binary mixes with size ratio u $\rightarrow \infty$, *i.e.* two noninteracting size classes, an analytical expression for the binary packing fraction is available [21, 39, 40]; the underlying concept and resulting expression are also applicable in $\mathbb{R}^D$ (D $\neq$ 3) (see Appendix B).

Revisiting and applying Onsager's excluded volume model to binary particle packings yields a generic and accurate packing expression applicable to the opposite limit: a small size difference (size ratios u from unity to at least two) applicable to RLP and RCP (and probably also to other intermediate states of packing), to (sphero)cylinders with all $\alpha$, and probably to many more particle shapes and other dimensions as well. This parameter-free closed-form expression is entirely characterized by the concentration (either number fraction $X_L$, Eq. (21) or volume fraction $c_L$, Eq. (23)) of the two components, particle size ratio u, and monosized packing fraction f. For the contraction function v(u), Eq. (34), based on [2], is appropriate.

Finally, it is noteworthy that the derived generic model is based solely on physical principles, and no adjustable parameters have been introduced to obtain the presented results. The governing parameters, *i.e.* the monosized packing fraction f, size ratio u, and concentration ($X_L$ or $c_L$), are all physically defined. For instance, the monosized packing fraction depends only on the considered particle shape (*e.g.* cylinder, spherocylinder, sphere) and the state of packing densification (*e.g.* loose, close).

## Appendix A Monosized packing of (sphero)cylinders

In this appendix, the packing fraction following from the excluded volume approach of Onsager [1] is compared with monodisperse packing data of spherocylinders and cylinders reported in the literature. The limiting behavior ($\alpha \downarrow 0$ and $\alpha \to \infty$) and the extrema of the monosized packing expressions for spherocylinders and cylinders are examined.

### A.1 Spherocylinders and cylinders, large l/d

For long spherocylinders and cylinders (having no end caps), one finds the same asymptotic relation because, for large l/d ($\alpha \downarrow 0$), the detailed shape of their ends is irrelevant. Furthermore, for small $\alpha$ (large l/d), random close and loose packing values both tend to zero (Section 6). First, reported packing values in this limit are presented.

Philipse [50] proposed a random contact model and related the total number of contacts per particle z and excluded volume to the packing fraction of (sphero)cylinders with a large aspect ratio l/d, arguing the packing fraction is proportional to the aspect ratio $\alpha$ (d/l). The number of contacts is probably the second most investigated property (after packing fraction) in studying particle packing, as it is a basic topological parameter that provides insight into packing configurations and their stability. Based on the contact model, the packing fraction of both cylinders and spherocylinders is asymptotically approximated by [50]

$$f = z\alpha\big(1 + O(\alpha)\big) \quad , \tag{A.1}$$

and z = 5.4 (± 0.2) was fitted to packing data. This $\alpha$ dependence also follows from Eqs. (6) and (7) for small $\alpha$. This equation is very similar to the equation ("Eq. (68)", rewritten here) introduced earlier by Pan [12]:

$$f = z\alpha\left(1 - \frac{z\pi\alpha}{2}\right)^{-1} = z\alpha\big(1 + O(\alpha)\big) \quad . \tag{A.2}$$

Blouwolff and Faden [61] counted the mean number of contacts per particle Z, *i.e.* the coordination number, in experimental random thin-rod packings and confirmed that Z is approximately 10 (Z = 2z) for long cylinders. In addition, they erroneously used the factor (2π + 3)/4 in Eq. (4), instead of the correct (π + 3)/4. This packing law, Eqs. (A.1)–(A.2), has been confirmed by experiments [10, 11, 62] as well as by simulations [63], yielding z in the range of 5.1-5.5.

**A.2 Spherocylinders, finite l/d**

Next, larger d/l are also considered, whereby the packing fraction spherocylinders and cylinders start deviating from each other. For spherocylinders, the packing fraction tends to a non-zero value for l/d tending to zero ($\alpha \rightarrow \infty$) (see Eq. (6)), that is, when the monosized packing of spheres is attained. This situation yields a packing fraction $f_{spe}$ = ¼ (see Eq. (6)), which does not match the established values of $f_{spe}^{rcp}$ nor $f_{spe}^{rlp}$ (Section 6). This value $f_{spe}$ = ¼ follows from the volume of 2 spheres ($\pi d^3/3$) and their excluded volume ($4\pi d^3/3$). A maximum packing fraction is observed for l/d ($\alpha^{-1}$) larger than zero [58, 63-65].

First, for spherocylinders Eq. (6) is generalized to

$$f_{spc}(\alpha) \approx z\left(\frac{\alpha+c_1\alpha^2}{1+c_2\alpha+c_3\alpha^2}\right) = z\left(\frac{\alpha^{-1}+c_1}{\alpha^{-2}+c_2\alpha^{-1}+c_3}\right) \quad . \tag{A.3}$$

Note that, for $\alpha \downarrow 0$, Eq. (A.1) is obtained. To the best of the author's knowledge, only random close packing of spherocylinders has been reported in the literature, and hence only this RCP configuration is addressed here.

| l/d | $f_{spc}^{rcp}$ | l/d | $f_{spc}^{rcp}$ | l/d | $f_{spc}^{rcp}$ |
|---|---|---|---|---|---|
| 0 | 0.645 | 0.7 | 0.675 | 1.8 | 0.633 |
| 0.1 | 0.672 | 0.8 | 0.673 | 2 | 0.629 |
| 0.2 | 0.685 | 0.9 | 0.668 | 2.2 | 0.614 |
| 0.3 | 0.689 | 1 | 0.659 | 2.5 | 0.604 |
| 0.35 | 0.686 | 1.2 | 0.656 | 3 | 0.576 |
| 0.4 | 0.687 | 1.4 | 0.650 | 3.5 | 0.558 |
| 0.5 | 0.684 | 1.5 | 0.643 | 4 | 0.536 |
| 0.6 | 0.679 | 1.6 | 0.642 | | |

**Table A.1** Computed RCP packing fraction values for spherocylinders for different l/d ($\alpha^{-1}$), taken from "Fig. 4" [58].

When the aspect ratio $\alpha \rightarrow \infty$ (so l/d $\downarrow$ 0), the RCP fraction of monosized spheres is obtained, $f_{spe}^{rcp}$, so Eq. (A.3) yields

$$c_3 = \frac{c_1 z}{f_{spe}^{rcp}} \quad , \tag{A.4}$$

From the literature, it is also known that that a maximum packing fraction is obtained for l/d ≈ 0.3-0.5, whereby $f_{spc,max}^{rcp} \approx 0.72$ [64, 65]. In [58, 63], a broad collection of published packing data on monosized spherocylinders for l/d up to 6 is reviewed, the data are consistent with this conclusion. Here, the computational data, extracted from [58, 63] and listed in Tables A.1 and A.2, are set out in Figure A.1. One can see the RCP sphere packing fraction that is attained at $\alpha^{-1}$ (= d/l) = 0, a maximum packing fraction at $\alpha_{max} \approx 2.6$, and a packing fraction that tends to zero for $\alpha \downarrow 0$ (l/d → ∞) (see previous subsection).

| **l/d** | $\mathbf{f_{spc}^{rcp}}$ |
|---|---|
| 0 | 0.634 |
| 0.4 | 0.695 |
| 1 | 0.682 |
| 2 | 0.615 |
| 4 | 0.536 |
| 8 | 0.420 |
| 20 | 0.226 |
| 40 | 0.128 |
| 80 | 0.060 |
| 120 | 0.034 |
| 160 | 0.025 |

**Table A.2** Computed RCP packing fraction values for spherocylinders for different l/d ($\alpha^{-1}$), derived from "Fig. 2" [63].

The aspect ratio α that yields this maximum packing fraction $\alpha_{max}$ follows from differentiating the right-hand side of Eq. (A.3) with respect to α and equating the result to zero, yielding

$$1 + 2\alpha_{max} + (c_1 c_2 - c_3)\alpha_{max}^2 = 0 \quad , \tag{A.5}$$

and the pertaining packing fraction is designated $f_{spc,max}^{rcp}$:

$$f_{spc,max}^{rcp} = z\left(\frac{\alpha_{max} + c_1\alpha_{max}^2}{1 + c_2\alpha_{max} + c_3\alpha_{max}^2}\right) \quad . \tag{A.6}$$

Combining Eqs. (A.4)-(A.6) now yields the following quadratic equation in $c_1$:

$$c_1^2\left(\frac{f^{rcp}_{spc,max}}{f^{rcp}_{spe}}-1\right)\alpha^3_{max}z + c_1\left(\left(\frac{f^{rcp}_{spc,max}}{f^{rcp}_{spe}}-1\right)\alpha^2_{max}z - f^{rcp}_{spc,max}\alpha_{max}\right) - f^{rcp}_{spc,max} = 0 \quad , \qquad (A.7)$$

yielding

$$c_1 = \frac{f^{rcp}_{spc,max}f^{rcp}_{spe}}{\left(f^{rcp}_{spc,max} - f^{rcp}_{spe}\right)\alpha^2_{max}z} \quad , \qquad (A.8)$$

$$c_2 = \frac{z}{f^{rcp}_{spc,max}} - \frac{2}{\alpha_{max}} \quad , \qquad (A.9)$$

and

$$c_3 = \frac{f^{rcp}_{spc,max}}{\left(f^{rcp}_{spc,max} - f^{rcp}_{spe}\right)\alpha^2_{max}} \qquad (A.10)$$

(see Eqs. (A.4) and (A.5)). With z = 5.3, $f^{rcp}_{spe}$ = 0.64, $\alpha_{max}$ = 2.6 (l/d = 0.385), and $f^{rcp}_{spc,max}$ = 0.72 [64, 65], it follows that $c_1$ = 0.161, $c_2$ = 6.592, and $c_3$ = 1.331. Eq. (A.6), with these values, is set out in Figure A.1.

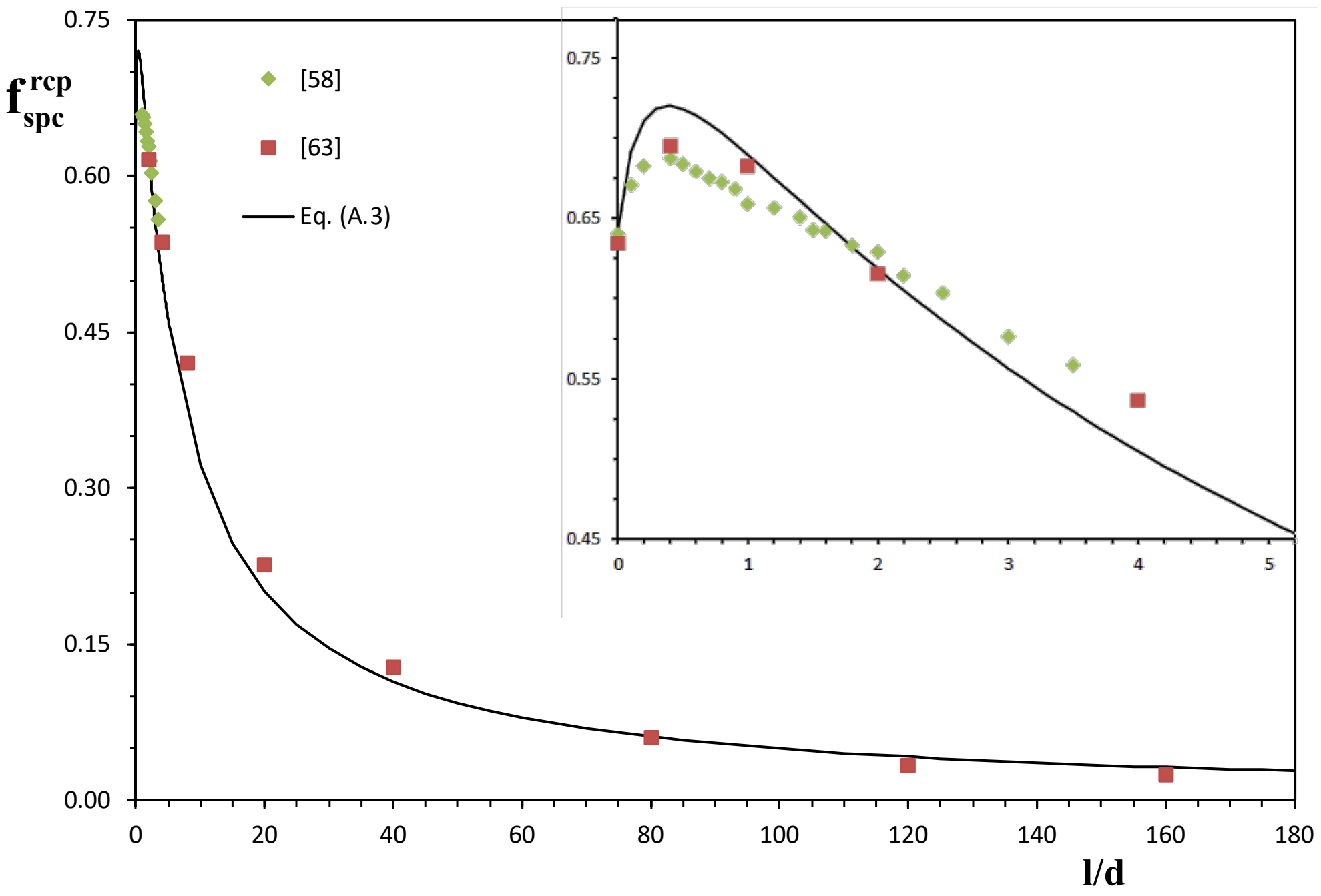

**Figure A.1** Random close packing fraction of monosized spherocylinders $f_{spc}^{rcp}$ as a function of the aspect ratio l/d ($\alpha^{-1}$) according to Eq. (A.6) with z = 5.3, $c_1 = 0.161$, $c_2 = 6.592$ and $c_3 = 1.331$. The computational data of [58, 63] are included as well (Tables A.1 and A.2). The inset shows a magnified view of the same graph for small l/d.

Figure A.1 shows good agreement between the fitted Onsager-based model, Eq. (A.6), and the spherocylinder random close packing values reported in the literature (Tables A.1 and A.2).

### A.3 Cylinders, finite l/d

In contrast to spherocylinders, Eq. (7) reveals that for cylinders the packing fraction tends to zero for both d/l ($\alpha$) and l/d ($\alpha^{-1}$) tending to zero, with a maximum monosized packing fraction for l/d ($\alpha^{-1}$) larger than zero, the same as for spherocylinders.

| **l/d** | $f_{cyl}^{rcp}$ | **l/d** | $f_{cyl}^{rcp}$ |
|---|---|---|---|
| 0.86 | 0.720 | 15.7 | 0.312 |
| 1.7 | 0.636 | 16.2 | 0.289 |
| 4.0 | 0.579 | 19.1 | 0.255 |
| 6.9 | 0.491 | 24.8 | 0.205 |
| 7.2 | 0.507 | 25.5 | 0.195 |
| 8.1 | 0.391 | 37.4 | 0.132 |
| 10.5 | 0.416 | 49.3 | 0.103 |
| 15.2 | 0.266 | 51 | 0.105 |

Table A.3 Experimental RCP packing fraction values for cylinders for different l/d ($\alpha^{-1}$), taken from "Fig. 4" [9].

This trend has been observed in literature too [60]. For cylinders, both random loose and random close packing experiments are reported in literature.

First, Eq. (7) is generalized to

$$f_{cyl}(\alpha) = z\left(\frac{\alpha}{1+c_1\alpha+c_2\alpha^2}\right) = z\left(\frac{\alpha^{-1}}{\alpha^{-2}+c_1\alpha^{-1}+c_2}\right) \quad . \tag{A.11}$$

| **l/d** | $f_{cyl}^{rcp}$ | **l/d** | $f_{cyl}^{rcp}$ |
|---|---|---|---|
| 0.53 | 0.617 | 60 | 0.081 |
| 0.63 | 0.613 | 90 | 0.050 |
| 1 | 0.629 | 34 | 0.197 |
| 1.2 | 0.607 | 49 | 0.080 |
| 1.3 | 0.585 | 59 | 0.070 |
| 11 | 0.403 | 60 | 0.087 |
| 15 | 0.300 | 90 | 0.056 |
| 34 | 0.145 | 167 | 0.033 |

**Table A.4** Experimental RCP packing fraction values for cylinders for different l/d ($\alpha^{-1}$), taken from "Tableau 4" and "Tableau 5" [11].

Note that, for $\alpha \downarrow 0$, Eq. (A.11) yields Eq. (A.1), and that for $\alpha \to \infty$ (so l/d $\downarrow$ 0), this equation yields a packing fraction tending to zero as well. It turs out that Eq. (A.11), when using the same $\alpha_{max}$ and $f_{cyl,max}^{rcp}$, just as for spherocylinders, good agreement is obtained.

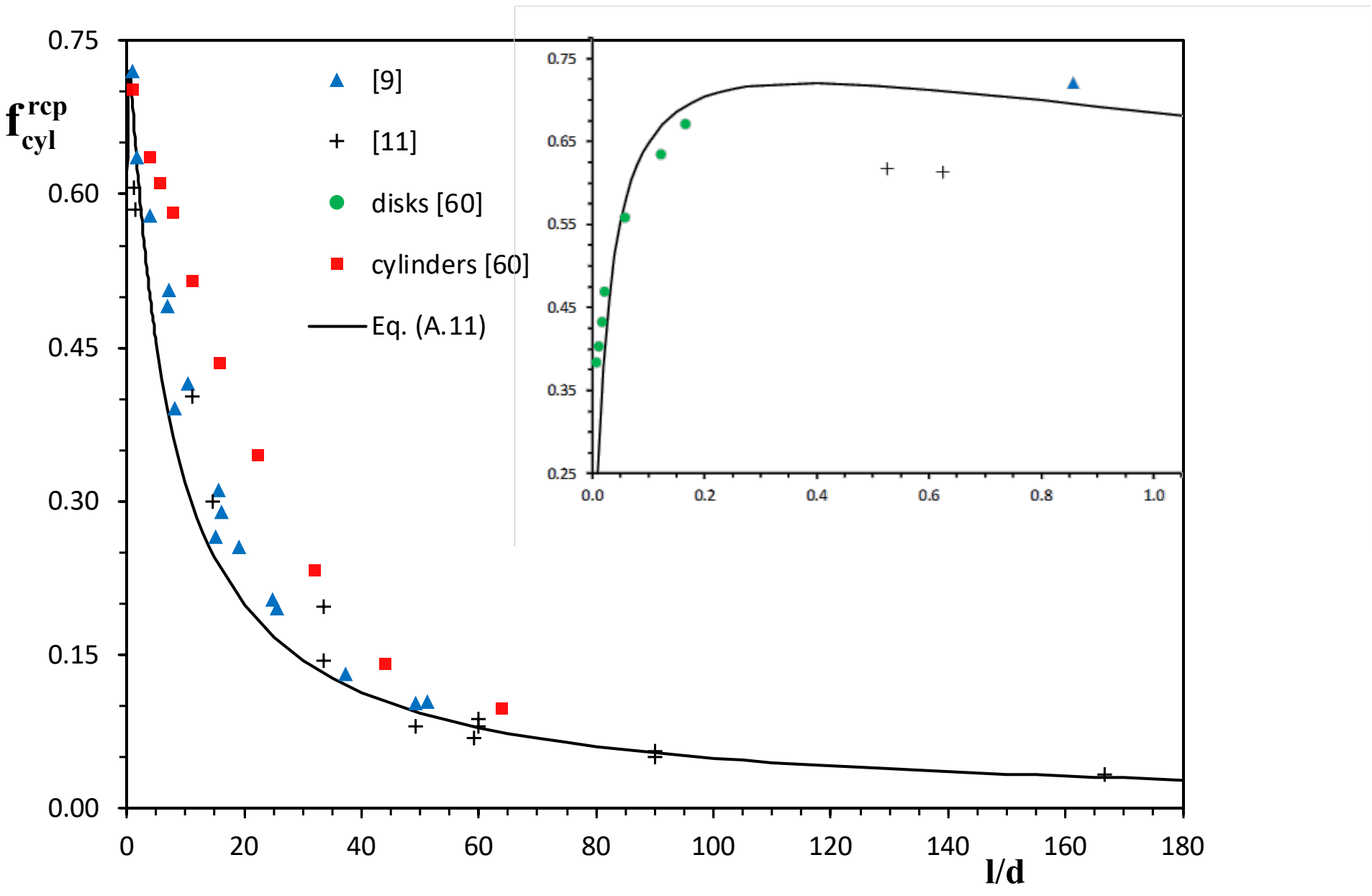

**Figure A.2** Random close packing fraction of monosized cylinders $f_{cyl}^{rcp}$ as a function of aspect ratio l/d ($\alpha^{-1}$) according to Eq. (A.11) with z = 5.3, $c_1$ = 6.592, and $c_2$ = 0.148. Experimental data from [9, 11, 60] are included as well (Tables 7, A.3, and A.4). Inset shows a magnified view of the same graph for small l/d.

The aspect ratio $\alpha$ that yields the maximum packing fraction, $\alpha_{max}$, follows from differentiating the right-hand side with respect to $\alpha$ and equating the result to zero, yielding

$$c_1 = \frac{z}{f_{cyl,max}} - \frac{2}{\alpha_{max}} \quad , \tag{A.12}$$

and the pertaining packing fraction is designated $f_{cyl,max}$, Eqs. (A.11) and (A.12) yielding

$$c_2 = \frac{1}{\alpha_{max}^2} \quad . \tag{A.13}$$

In Figure A.2, measured random close packing data for cylinders and disks are included [9, 11, 60], listed in Tables 7, A.3, and A.4. Onsager-based Eq. (A.11) is fitted using $c_1 = 6.592$ and $c_2 = 0.148$, that follow from $z = 5.3$, $\alpha_{max} = 2.6$, and $f_{cyl,max}^{rcp} = 0.72$ (see Eqs. ((A.12) and (A.13)).

| l/d | $f_{cyl}^{rlp}$ |
|---|---|
| 0.53 | 0.556 |
| 0.63 | 0.546 |
| 1 | 0.538 |
| 1.2 | 0.548 |
| 1.3 | 0.505 |

**Table A.5** Experimental RLP packing fraction values for cylinders for different l/d ($\alpha^{-1}$), taken from "Tableau 5" [11].

In Figure A.3, measured random loose packing data for cylinders and disks are included [11, 60], listed in Tables 7 and A.5. Onsager-based Eq. (A.11) is fitted using $z = 3.3$, *i.e.* fewer contacts per particle than for RCP, $\alpha_{max} = 2.6$ and $f_{cyl,max}^{rlp} = 0.58$, and from Eqs. (A.12) and (A.13) it follows that $c_1 = 4.920$ and $c_2 = 0.148$, respectively.

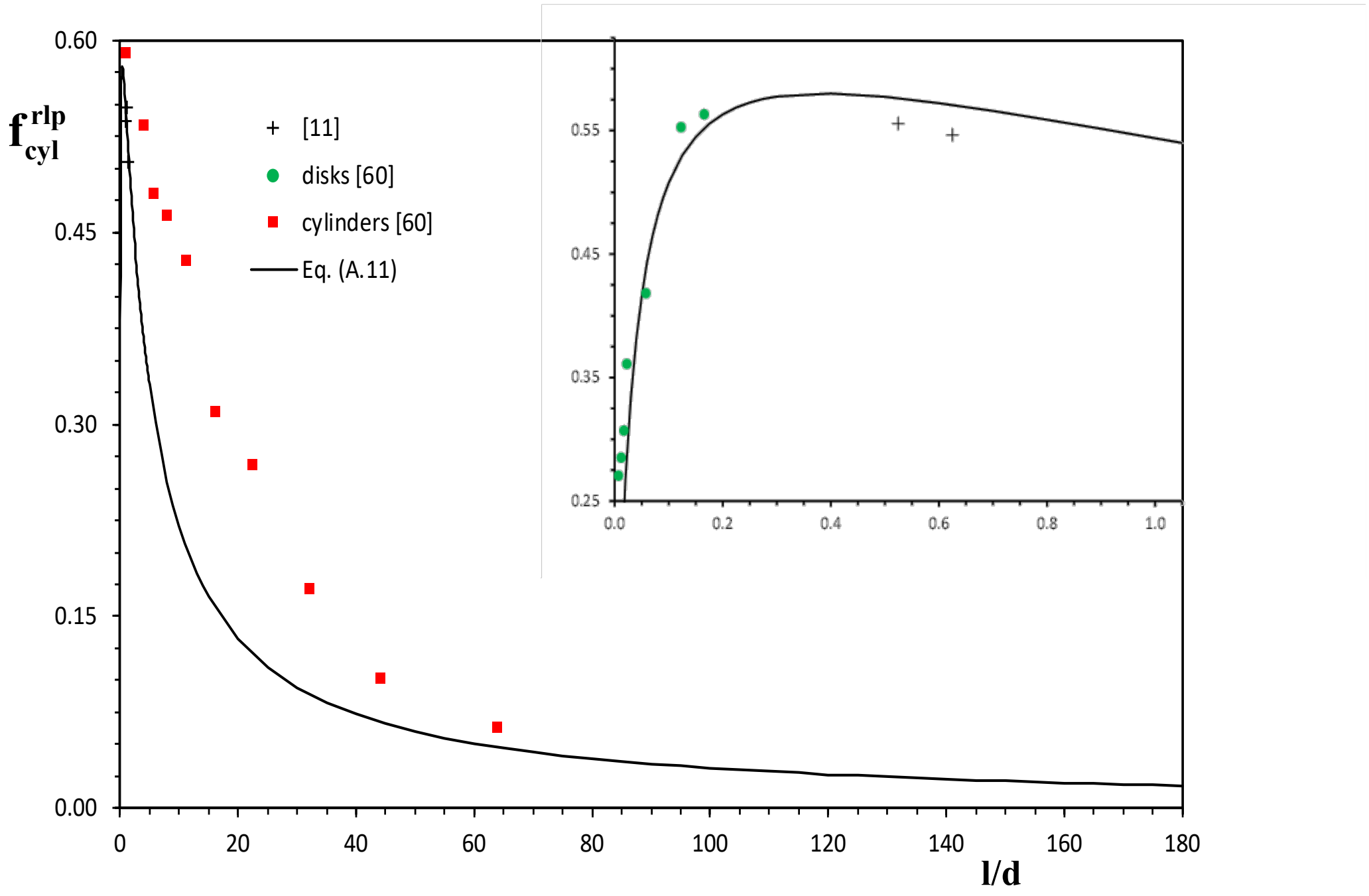

**Figure A.3** Random loose packing fraction of monosized cylinders $f_{cyl}^{rlp}$ as a function of the aspect ratio l/d ($\alpha^{-1}$) according to Eq. (A.11) with $z = 3.3$, $c_1 = 4.920$, and $c_2 = 0.148$. Experimental data from [11, 60] are included as well (Tables 7 and A.5). Inset shows a magnified view of the same graph for small l/d.

Also here, we see that the Onsager-based model captures the packing fractions quite well in the entire $\alpha$ (or l/d) range.

In this appendix, modified Onsager-based equations have been applied to RCP of (sphero)cylinders and RLP of cylinders, based on a thorough study of reported computational and experimental data. Importantly, for all three monodisperse assemblies, the maximum packing seems to take place for aspect ratio $\alpha_{max} \approx 2.6$ (l/d $\approx$ 0.38). Furthermore, the maximum RCP packing fraction of spherocylinders and cylinders seems to be very similar, $f^{rcp}_{spc,max} \approx f^{rcp}_{cyl,max} \approx 0.72$. Also, for large l/d, spherocylinders and cylinders feature the same RCP packing fraction (governed by contact number z).

## Appendix B Binary packing fraction of similar particles in $\mathbb{R}^D$ ($D \neq 3$)

In this appendix, the methodology that has resulted in the binary packing fraction expression in 3-dimensional Euclidean space (D = 3), namely Eq. (21), is here employed to derive an expression for the packing of hard cylinders (disks) in $\mathbb{R}^2$ and hard hyperspheres ("D-spheres") in $\mathbb{R}^D$ for D > 3, with D as the dimensionality. The optimal packing of hard spheres in higher dimensions is, for instance, of interest in error-correcting codes in communication theory [3]. Also, here polydispersity by combining two sizes will result in higher packing fractions. As D ≠ 3 is discussed in this appendix, a subscript "D" is added to monosized packing fraction f, binary packing fraction η, and contraction function v, so that f, η and v are henceforth termed $f_3$, $\eta_3$, and $v_3$, respectively.

### B.1 Packing of binary disks/cylinders in a plane

First, the packing of binary cylinders (or disks or circles) in a plane will be discussed, using the excluded volume approach of Onsager used for (sphero)cylinders in Section 2. The convex combination of the surface area of two cylinders with diameters $d_1$ and $d_2$ and number fractions $X_1$ and $X_2$, respectively, reads

$$2A_{cyl} = \tfrac{\pi}{2}(X_1 d_1^2 + X_2 d_2^2) \quad , \tag{B.1}$$

and the excluded volume of the two cylinders (see Figure 1, which is also applicable to cylinders in $\mathbb{R}^2$) is

$$A_e^{1,2} = \tfrac{\pi}{4}(d_1 + d_2)^2 \quad . \tag{B.2}$$

The statistically averaged excluded surface area of randomly mixed cylinders follows from

$$A_e = \sum_{i=1}^{2}\sum_{j=1}^{2} X_i X_j A_e^{i,j} = X_1 A_e^{1,1} + X_2 A_e^{2,2} - X_1 X_2 (A_e^{1,1} + A_e^{2,2} - 2A_e^{1,2}) \quad , \tag{B.3}$$

whereby Eq. (10) and

$$A_e^{1,2} = A_e^{2,1} \tag{B.4}$$

have been employed. Inserting Eq. (B.2) yields

$$A_e = \pi\left(X_1 d_1^2 + X_2 d_2^2 - X_1 X_2 (d_1^2 + d_2^2 - \tfrac{1}{2}(d_1 + d_2)^2\right) =$$

$$\pi\left(X_1 d_1^2 + X_2 d_2^2 - \tfrac{1}{2} X_1 X_2 (d_1 - d_2)^2\right) \quad . \tag{B.5}$$

The binary packing density for D = 2 is obtained by dividing Eq. (B.1) by Eq. (B.5), producing

$$\eta_2 = \frac{X_1 d_1^2 + X_2 d_2^2}{2\left(X_1 d_1^2 + X_2 d_2^2 - \tfrac{1}{2} X_1 X_2 (d_1 - d_2)^2\right)} \quad . \tag{B.6}$$

For $X_1 = 0$ ($X_2 = 1$), $X_1 = 1$ ($X_2 = 0$), or $d_1 = d_2$, the monosized packing fraction $f_2 = ½$ follows from using Onsager's excluded volume. From the analysis of the 3-dimensional packing of cylinders and spherocylinders with arbitrary l/d, we presume that the factor ½ in the denominator is not $f_2$, but $1 – f_2$. Furthermore, for this monodisperse packing fraction, the true monosized packing fraction needs to be substituted. This all also held for the Onsager excluded volume-based factor $1 – f_3$ in the contraction term (Section 4). In recent papers, values for $f_2$ are recapitulated, depending on compaction (from RLP to RCP) they range from 0.66 to 0.89 [26, 66].

Applying Eqs. (10) and (11) to Eq. (B.6) and setting $X_L = X_1$ now results in

$$\eta_2(u, X_L) = \frac{f_2 (X_L(u^2 - 1) + 1)}{X_L(u^2 - 1) + 1 - X_L(1 - X_L)(1 - f_2)v_2(u)} \tag{B.7}$$

with as contraction function

$$v_2(u) = (u - 1)^2 \quad . \tag{B.8}$$

Analogous to Eq. (21) for (sphero)cylinders in D = 3, using Onsager's excluded volume leads to a similar expression for the packing fraction of cylinders (disks) in a plane.

As with the packing of (sphero)cylinders in $\mathbf{R}^3$, Eq. (B.7) contains a factor $v_2(u)$ that is quadratic in (u – 1), originating from the $(d_1 - d_2)^2$ factor in Eq. (B.6). This factor confirms that the contraction does not depend on the sign of $d_1 - d_2$, in other words, on whether $d_1 > d_2$ or $d_1 < d_2$. The empirical contraction provided by [2], Eq. (25), which is quadratic $(V_{1,spe} - V_{2,spe})^2$, features the same characteristic. But the contraction function of the binary expression of spheres, Eq. (22), also contains the factor (u + 1), which is absent in Eq. (B.8).

Apparently, the contraction function that results from following the excluded volume approach is the product of $(u - 1)^2$ and a function that depends on the dimension D. Accordingly, in the spirit of the D = 3 case, the following general contraction function is introduced:

$$v_D(u) = w_D(u)\,(u-1)^2 \quad , \tag{B.9}$$

whereby $w_2(u) = 1$ (Eq. (B.8)), and $w_3(u) = u + 1$ ((Eq. (22)).

In Section 4, it was concluded that using Eq. (22) in Eq. (21) provided results that were in line with simulations as long as $u - 1 \approx 0$, but that, for larger size ratios, Eq. (34) provided better agreement. For u = 2, Eq. (22) yields $v_3(u = 2) = 3$, whereas Eq. (34) yields $v_3(u = 2) = 2.42$. The comparison shown in Figure 3 confirms that Eq. (22) overestimates the binary packing fraction by about 20%. Translating Eq. (34) to the $\mathbf{R}^2$ case suggests the following alternative contraction function:

$$v_2(u) = \frac{(u^2-1)^2}{2(u^2+1)} = \frac{(u+1)^2(u-1)^2}{2(u^2+1)} \quad . \tag{B.10}$$

This expression is asymptotically identical to Eq. (B.8) for u ↓ 1 but might be more accurate for larger u. For u = 2, Eq. (B.8) yields $v_2(u = 2) = 1$, whereas Eq. (B.10) yields the slightly lower value $v_2(u = 2) = 0.9$.

**B.2 Packing of hyperspheres in D-dimensional (D > 3) space**

Here, we will study the packing of binary hyperspheres following the approach of the previous subsection. The convex volume addition of two binary hyperspheres with diameters $d_1$ and $d_2$ and number fractions $X_1$ and $X_2$, respectively, reads [67];

$$2V_{spe} = \frac{4\left(\frac{\sqrt{\pi}}{2}\right)^D (X_1 d_1^D + X_2 d_2^D)}{D\,\Gamma\left(\frac{D}{2}\right)} \quad , \qquad \text{(B.11)}$$

with Γ as the gamma function. Eq. (B.11) yields Eqs. (1) and (B.1) for D = 3 (l/d = 0) and D = 2, respectively. The excluded volume of these two hyperspheres is

$$V_e^{1,2} = \frac{2\left(\frac{\sqrt{\pi}}{2}\right)^D (d_1 + d_2)^D}{D\,\Gamma\left(\frac{D}{2}\right)} \quad . \qquad \text{(B.12)}$$

Eqs. (15), (B.11), and (B.12) yield the bidisperse packing fraction

$$\eta_D = \frac{2(X_1 d_1^D + X_2 d_2^D)}{2^D X_1 d_1^D + 2^D X_2 d_2^D - X_1 X_2 (2^D d_1^D + 2^D d_2^D - 2(d_1 + d_2)^D)} \quad . \qquad \text{(B.13)}$$

For $X_1 = 0$ ($X_2 = 1$), $X_1 = 1$ ($X_2 = 0$), or $d_1 = d_2$, the monosized packing fraction

$$f_D = 2^{1-D} \qquad \text{(B.14)}$$

follows. This power-law scaling with the dimension D is in qualitative agreement with $f_D = 2^{1-D}$ $(0.023\ D^2 + 0.61\ D + 0.365)$ [51], $f_D \sim 2^{-D}(D \log D)$ [68] and $f_D = 2^{1-D}(1.28\ D - 1.36)$ [69]. Note that the first expression is based on applying Onsager's excluded volume to monodisperse hyperspheres [51].

Invoking $2^{1-D} = f_D$ in the nominator and introducing Eqs. (10) and (11), u > 1, and $X_1 = X_L$ yields

$$\eta_D(u, X_L) = \frac{f_D(X_L(u^D - 1) + 1)}{X_L(u^D - 1) + 1 - X_L(1 - X_L)(u^D + 1 - 2^{1-D}(u+1)^D)} \quad . \qquad \text{(B.15)}$$

With the derivation of the $\mathbb{R}^3$ binary packing of spherocylinders and cylinders (Section 2), the factor 1 - f (here designated as $1 - f_3$) appeared in the contraction term (see Eqs (18) and (20), respectively). We assume this to be applicable to all D, as was done in Subsection B.1 for $\mathbb{R}^2$. This implies that we write

$$u^D + 1 - 2^{1-D}(u + 1)^D = (1 - 2^{1-D})w_D(u)\,(u - 1)^2 \quad . \qquad \text{(B.16)}$$

For D = 2 and D = 3, this equation yields the known $w_2(u) = 1$, and $w_3(u) = u + 1$. The conclusion concerning an excluded volume analysis of 3-dimensional spheres might only be that the contraction term is $(u^D + 1) – f_D(u + 1)^D$, *i.e.* the left-hand side of Eq. (B.16) with Eq. (B.14) substituted. Here, the contraction term will be defined by the product of $(1 – f_D)$ and Eq. (B.9), whereby $w_D(u)$ follows from Eq. (B.16). Namely from spherocylinders and cylinders, we learned in Section 2 (Eqs. (18) and (20)) that, in the particular case of spheres, it holds that $f_3$ = ¼, and $u^3 + 1 – (u + 1)^3/4$ is written as the product $(1 – 1/4)(u + 1)(u – 1)^2 = (1 – f_3)v_3(u) = (1 – f_3)\, w_3(u)\, (u – 1)^2$, and not as $u^3 + 1 – (u + 1)^3/4 = u^3 + 1 – (u + 1)^3 f_3$. It is tentatively conjectured that this conclusion also holds for hyperspheres (D > 3).

Based on this principle, for D > 3, the expression for $w_D(u)$ can be derived from Eq. (B.16), and are listed in Table B1. As a result, for all D, the contraction functions $v_D(u)$ are the product of the factors $(u – 1)^2$ and $w_D(u)$, the latter depending on dimension D, the result listed in Table B.1.

| D | $w_D(u)$ | $w_D(u = 1)$ |
|---|---|---|
| 2 | 1 | 1 |
| 3 | $u + 1$ | 2 |
| 4 | $u^2 + 10u/7 + 1$ | 24/7 |
| 5 | $u^3 + 5u^2/3 + 5u/3 + 1$ | 16/3 |
| 6 | $u^4 + 56u^3/31 + 66u^2/31 + 56u/31 + 1$ | 240/31 |
| 7 | $u^5 + 17u^4/9 + 22u^3/9 + 22u^2/9 + 17u/9 + 1$ | 32/3 |
| 8 | $u^6 + 246u^5/127 + 337u^4/127 + 372u^3/127 + 337u^2/127 + 246u/127 + 1$ | 1792/127 |
| 9 | $u^7 + 501u^6/255 + 711u^5/255 + 837u^4/255 + 837u^3/255 + 711u^2/255 + \ldots$.. | 1536/85 |
| 10 | $u^8 + 1012u^7/511 + 1468u^6/511 + 1804u^5/511 + 1930u^4/511 + 1804u^3/511 + \ldots$. | 11394/511 |

**Table B.1** Function $w_D(u)$ for various dimensions D that follows from the excluded volume approach and writing resulting contraction function, Eq. (B.16), in the form of Eq. (B.9). Due to the symmetry of all polynomials, expressions for D = 9 and 10 are truncated.

Hence, the binary packing fraction in D-dimensional space reads

$$\eta_D(u, X_L) = \frac{f_D(X_L(u^D - 1) + 1)}{X_L(u^D - 1) + 1 - X_L(1 - X_L)(1 - f_D)v_D(u)} \quad , \qquad \text{(B.17)}$$

with $v_D(u)$ following from Eq. (B.9), and $w_D(u)$, from Table B.1, $f_D$ may be taken from [51, 68, 69].

As with the 2- and 3-dimensional spaces, also for D > 3, a Mangelsdorf and Washington-based contraction function is proposed which might be more accurate for larger size ratios:

$$v_D(u) = w_D(u=1)\frac{2(u^D - 1)^2}{D^2(u^D + 1)} \quad , \tag{B.18}$$

where $w_D(u = 1)$ can be found in Table B.1; and the values are depicted in Figure B.1, for D ranging from 2 to 10.

Eq. (B.18) tends asymptotically to Eq. (B.9) for $u \downarrow 1$ and is proportional to $u^D$ for larger u, as is also the case with Eq. (B.9). From Table B.1, one can see that $w_D$ is proportional to $u^{D-2}$, and hence $v_D$ to $u^D$ (Eq. (B.9)). Obviously, for D = 3 and D = 2, Eq. (B.18) yields Eqs. (34) and (B.10), respectively. And, as with 2- and 3-dimensional spaces, Eq. (B.18) provides a smaller contraction value than the Onsager-based contraction function. For instance, for D = 10 and u = 2, the contraction value $v_{10}(u = 2)$ computed from Eq. (B.18) and $w_{10}(u = 1)$ from Table B.1 is about half of that when computing it with Eq. (B.9) and taking $w_{10}(u = 2)$ from the expression given in Table B.1.

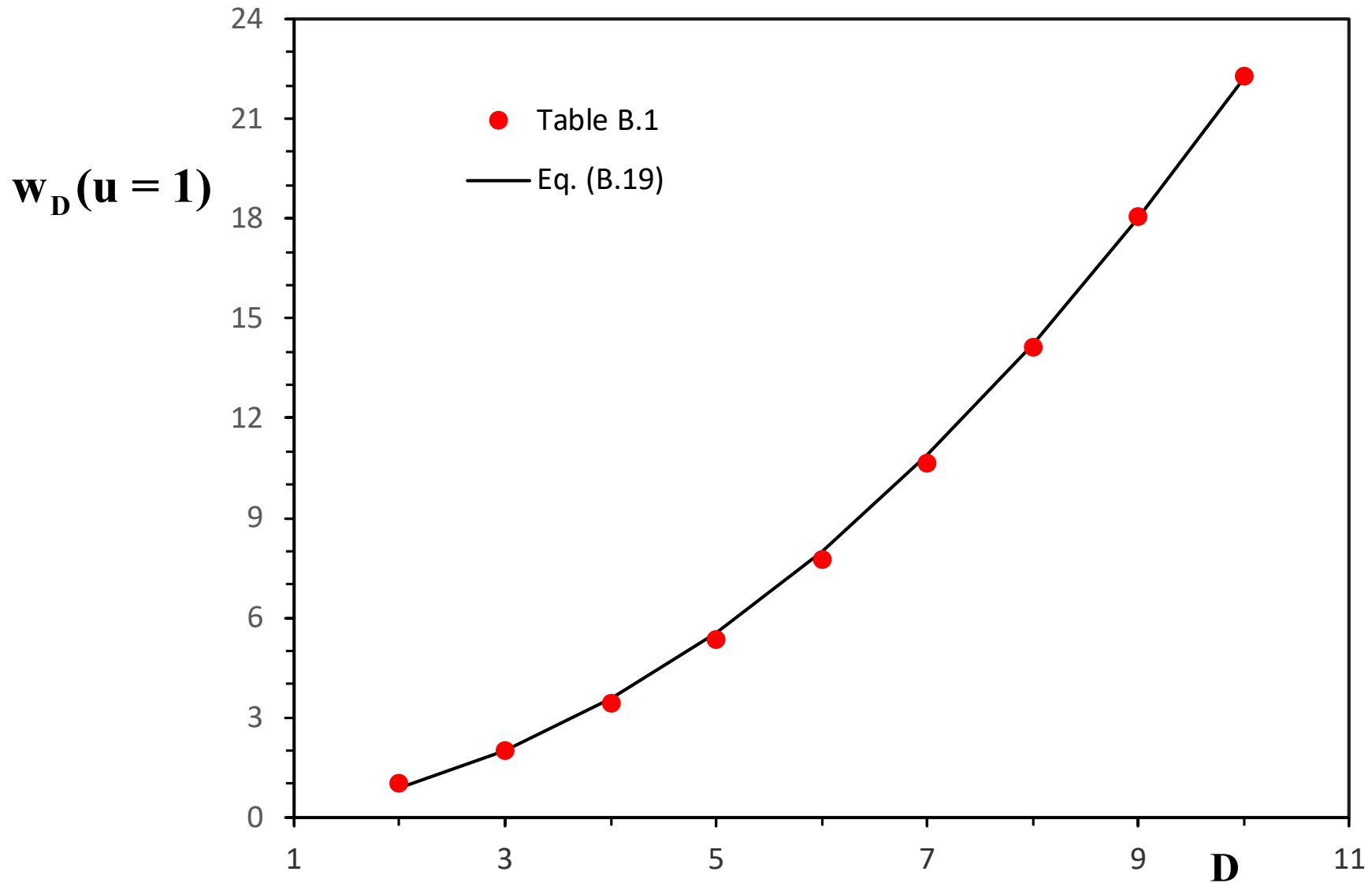

**Figure B.1** Values of $w_D(u = 1)$ as a function of dimension D, taken from Table B.1 and computed with Eq. (B.19).

For D = 2,…, 10, the values of $w_D(u = 1)$ can be found in Table B.1. For larger space dimensions D, the value can be assessed by the following fit:

$$w_D(u=1) = \frac{2}{9} D^2 \quad , \tag{B.19}$$

which is included in Figure B.1. This fit of $w_D(u = 1)$ values is least accurate for D = 2, exact for D = 3, and remains accurate at higher dimensionalities D (Table B.1). It is striking that $w_D(u = 1)$ is proportional to $D^2$, which cancels out with the factor $D^2$ in the denominator of Eq. (B.18). In this appendix, a method is proposed to predict the packing fraction of similar particles with a small size difference, so u close to unity, in D-dimensional space $\mathbb{R}^D$ ($D \neq 3$). In this derivation, the insights gathered with binary spherocylinders and cylinders in the 3-dimensional space are applied. However, the applicability of Eq. (B.17), with either Eq. (B.9) or Eq. (B.18), still needs to be confirmed by experiments (D = 2) and/or simulations (D = 2, D > 3). Notably, whereas the monosized packing fraction *decreases* with dimension D (Eq. (B.14), [53], [71], [72]), the packing fraction enlargement by dispersity *increases* with D (Eqs. (B.16) and (B.18)).

For binary mixes with a large size ratio u ($u \to \infty$) in $\mathbb{R}^3$, *i.e.* two noninteracting fractions, being the opposite limit of $u \downarrow 1$, analytical expressions for the binary packing fraction are also available [21, 39, 40]. Notably, the underlying principle of non-interacting similar particles is applicable to all $\mathbb{R}^D$, including $D \neq 3$. In all $\mathbb{R}^D$ for $u \to \infty$, namely smaller particles with packing fraction $f_D$ are able to fill the voids of the large particles that have the void fraction $(1 - f_D)$.

**Acknowledgements**

The author wishes to thank I. Biazzo, F. Zamponi, M. Clusel, J. Brujić, M. Danisch, H.A. Makse, K.W. Desmond, and E. Weeks for providing computer simulations of binary sphere packings. He furthermore thanks A. Kaja and V. Elfmarkova for the data extracted from "Fig. 10a" [58] and "Fig. 2 "[60], respectively, and Y. Luo for the data from "Fig. 4" [9], "Fig. 4" [58], and "Fig. 2" [63].